\documentclass[12pt]{article}

\usepackage{amsfonts,amsmath,amssymb,accents,mathrsfs}
\usepackage[retainorgcmds]{IEEEtrantools}
\usepackage{multirow}
\usepackage{multicol}
\usepackage{tikz}
\usepackage{graphicx,color}
\usepackage{booktabs}
\usepackage{placeins}
\usepackage[colorlinks,linkcolor=Blue,citecolor=Blue,urlcolor=Blue,bookmarks,bookmarksnumbered]{hyperref}
\usepackage[nosort]{cite}
\usepackage[scaled=0.85]{helvet}
\usepackage[utf8]{inputenc}
\usepackage{cancel}

\definecolor{Red}    {rgb}{0.90,0.00,0.12} %  1
\definecolor{Blue}   {rgb}{0.00,0.00,1.00} %  2
\definecolor{Green}  {rgb}{0.10,0.70,0.10} %  3
\definecolor{Turque} {rgb}{0.00,0.65,0.85} %  4
\definecolor{Orange} {rgb}{1.00,0.50,0.15} %  5
\definecolor{Magenta}{rgb}{1.00,0.00,1.00} %  6
\definecolor{Gold}   {rgb}{1.00,0.75,0.25} %  7
\definecolor{Seaweed}{rgb}{0.01,0.24,0.09} %  8
\definecolor{Purple} {rgb}{0.50,0.25,0.55} %  9
\definecolor{Brown}  {rgb}{0.43,0.26,0.32} % 10
\definecolor{grey1}  {rgb}{0.20,0.20,0.20} % 11
\definecolor{grey2}  {rgb}{0.40,0.40,0.40} % 12
\definecolor{grey3}  {rgb}{0.60,0.60,0.60} % 13
\definecolor{grey4}  {rgb}{0.80,0.80,0.80} % 14
\definecolor{grey5}  {rgb}{0.90,0.90,0.90} % 15

\def\a{{\alpha}}
\def\b{{\beta}}
\def\g{{\gamma}}
\def\d{{\delta}}
\def\r{{\rho}}
\def\s{{\sigma}}

\def\L{{\Lambda}}

\def\ad{{\dot{\alpha}}}
\def\bd{{\dot{\beta}}}

\def\N{{\mathcal{N}}}

\def\J{{\mathcal{J}}}
\def\T{{\mathcal{T}}}

\def\K{{\mathcal{K}}}
\def\W{{\mathcal{W}}}
\def\F{{\mathcal{F}}}

\def\D{{\rm D}}
\def\Dd{{\bar{\rm D}}}
\def\pa{\partial}

\def\be{\begin{equation}}
\def\ee{\end{equation}}
\def\bea{\begin{IEEEeqnarray*}}
\def\eea{\end{IEEEeqnarray*}}
\def\n{\IEEEyesnumber}
\def\sn{\IEEEyessubnumber}

\makeatletter
\def\section{\@startsection{section}{1}{\z@}
              {3ex plus-1ex minus-.2ex}{1pt plus1pt}{\large\sf\bfseries\boldmath}}
\def\subsection{\@startsection{subsection}{2}{\z@}
              {1.5ex plus-1ex minus-.2ex}{0.01pt plus1pt}{\sf\slshape}}
\def\subsubsection{\@startsection{subsubsection}{3}{\z@}
              {1.5ex plus-1ex minus-.2ex}{0.01pt plus0.2pt}{\sf\boldmath}}
\def\paragraph{\@startsection{paragraph}{4}{\z@}
              {.75ex \@plus.5ex \@minus.2ex}{-2mm}{\sf\bfseries\boldmath}}
\makeatother

   \parskip\medskipamount     \lineskip=0pt
\thicklines
\begin{document}
\thispagestyle{empty}
\noindent{\small
\hfill{HET-1765 {~} \\ % un-comment-out and specify when done}
$~~~~~~~~~~~~~~~~~~~~~~~~~~~~~~~~~~~~~~~~~~~~~~~~~~~~~~~~~~~~$
$~~~~~~~~~~~~~~~~~~~~\,~~~~~~~~~~~~~~~~~~~~~~~~~\,~~~~~~~~~~~~~~~~$
 {~}
}
\vspace*{6mm}
\begin{center}
{\large \bf Interaction of supersymmetric nonlinear sigma models
with external higher spin superfields via higher
spin supercurrents \vspace{3ex}
%Constraining interactions of supersymmetric sigma models\\
%with massless higher spin supermultiplets\vspace{3ex}\\
%
%[Constraining supersymmetric sigma models\\
%with higher spin supercurrents]
} \\   [9mm] {\large { I. L.
Buchbinder,\footnote{joseph@tspu.edu.ru}$^{a,b,c}$ S.\ James Gates,
Jr.,\footnote{sylvester${}_-$gates@brown.edu}$^{d}$ and Konstantinos
Koutrolikos\footnote{kkoutrolikos@physics.muni.cz}$^{e}$ }}
\\*[8mm]
\emph{
\centering
$^a$Department of Theoretical Physics,Tomsk State Pedagogical University,\\
Tomsk 634041, Russia
\\[6pt]
$^b$National Research Tomsk State University,\\
Tomsk 634050, Russia
\\[6pt]
$^c$Departamento de F\'isica, ICE, Universidade Federal de Juiz de Fora,\\
Campus Universit\'ario-Juiz de Fora, 36036-900, MG, Brazil
\\[6pt]
$^{d}$Department of Physics, Brown University,
\\[1pt]
Box 1843, 182 Hope Street, Barus \& Holley 545,
Providence, RI 02912, USA
\\[6pt]
$^e$ Institute for Theoretical Physics and Astrophysics, Masaryk
University,
\\[1pt]
611 37 Brno, Czech Republic
}
 $$~~$$
  $$~~$$
 \\*[-8mm]
{ ABSTRACT}\\[4mm]
\parbox{142mm}{\parindent=2pc\indent\baselineskip=14pt plus1pt
We consider a four dimensional generalized Wess-Zumino model
formulated in terms of an arbitrary K\"{a}hler potential
$\K(\Phi,\bar{\Phi})$ and an arbitrary chiral superpotential
$\W(\Phi)$. A general analysis is given
to describe the possible interactions of this theory
with external higher spin gauge superfields of the ($s+1,s+1/2$)
supermultiplet via higher spin
supercurrents. It is shown that such interactions do not exist beyond
supergravity
$(s\geq2)$ for any $\K$ and $\W$. However, we find three exceptions,
the theory of a free massless chiral, the theory of a free massive
chiral and the theory of a free chiral with linear superpotential.
For the first two, the higher spin supercurrents are known and for
the third one we provide the explicit expressions. We also discuss
the lower spin supercurrents. As expected, a coupling to
(non-minimal) supergravity ($s=1$) can always be found and we give
the generating supercurrent and supertrace for arbitrary $\K$ and
$\W$. On the other hand, coupling to the vector supermultiplet
($s=0$) is possible only if $\K=\K(\bar{\Phi}\Phi)$ and $\W=0$.}
\end{center}
$$~~$$
\vfill
\vfill
\clearpage
%
%%%%%%%%%%%%%%%%%%%%%%%%%%%%%%%%%%%%%%%%%%%%%%
\section{Introduction}
\label{sec:intro}
%%%%%%%%%%%%%%%%%%%%%%%%%%%%%%%%%%%%%%%%%%%%%%
Higher spin fields and their interactions are the subjects of
extensive study. Despite the various no-go theorems
\cite{nogo1,nogo2,nogo3,nogo4,nogo5,nogo6,nogo7,nogo8,
nogo9,nogo10,nogo11,nogo12,yesgo1,yesgo2,review1,review2}
and great efforts it is not clear yet whether higher spin fields
play a role in the description of fundamental physical
phenomena. Nevertheless, higher spin fields attract much attention
due to many remarkable features, e.g. their contribution in
the softness of string interactions by regularizing the ultraviolet
with an infinite tower of massive states and providing a framework
for studying and understanding the \emph{AdS/CFT} correspondence.
Furthermore, studying higher spin fields allows us to better
understand the structure of interactions in general gauge theory. In
many cases, interaction terms for higher spins were
successfully constructed for flat spacetime at first order in
coupling constant $g$ by using a variety of techniques, such as
light-cone approach \cite{lc1,lc2,lc3,lc4,lc5,lc6,lc7,lc8,lc9,lc10},
Noether's procedure \cite{Nm1,Nm2,Nm3,Nm4,Nm5} (some of these
results were later generalized in \cite{Gen0,Gen1,Gen2,Gen3}) and
BRST \cite{BRST1,BRST2,BRST3,BRST4,BRST5,BRST6,BRST7,BRST8}. In an
intriguing manner, most of the previously mentioned results together
with some new interaction vertices have been obtained by analyzing
tree level amplitudes of (super)strings
\cite{String1,String2,String3}, thus enhancing the connection
between string theory and higher spin fields. For (A)dS backgrounds
similar results have been obtained \cite{AdS1,AdS2,AdS3,AdS4,AdS5}
which eventually led to the fully interacting equations of motion
for higher spin fields \cite{fihs}.

Among these interactions, the simplest class is provided by the
cubic coupling of  higher spin fields with low spin matter fields,
such as scalar and spinor fields \cite{Nm3,BRST7,Scalar1,Scalar2}
(and \cite{Superspace1,Superspace2,
Superspace3,Superspace4,Superspace5} for supersymmetric
generalizations) which are of the type \emph{higher spin gauge field
$\times$ conserved current}, where the conserved current is quadratic
in the derivatives of the matter fields. The cubic nature of these
interactions is a consequence of the fact that we couple
non-interacting (free) matter fields to higher spins. It is natural
to take the next step and consider the coupling of interacting low
spin fields to external higher spin gauge field. In this paper we
ask this question and investigate the possibility of such 
interactions.

%However this is not a choice we make, it is enforced by the
%Coleman-Mandula theorem \cite{nogo3} (and its supersymmetric
%generalization \cite{nogo4}). If we consider interacting matter
%fields leading to a non-trivial $\mathcal{S}$-matrix the
%Coleman-Mandula theorem does not allow their coupling to higher
%spins. However, the Coleman-Mandula theorem can be bypassed in
%theories which do not have a conventional $\mathcal{S}$-matrix. For
%example, in
% \cite{Maldacena} this was explored for three dimensional conformal field
%theories and an extension of the Coleman-Mandula theorem was proven.

%The conventional definition of the scattering matrix is based on the
%existence of asymptotically free \emph{in} and \emph{out} states.
%Therefore, one can think that a complicated, non-linear theory with
%strong interactions may disrupt or not allow the definition of the
%free \emph{in} and \emph{out} states. Hence in that theory the
%Coleman-Mandula theorem does not hold any more, thus there might be
%non-cubic first order (in coupling constant $g$) interactions of the
%matter fields with higher spins.

In order to include both scalar and spinorial matter fields in our
discussion and simplify the technical details imposed by 
supersymmetry we will consider a theory of the chiral
supermultiplet described by a chiral superfield $\Phi$.
For its dynamics we will assume a nonlinear
supersymmetric sigma model described by an arbitrary K\"{a}hler
potential $\K(\Phi,\bar{\Phi})$ with the addition of an arbitrary
chiral superpotential $\W(\Phi)$ (see e.g.\cite{BK}). Such a model
is a good parametrization of many interacting matter theories and a
good candidate for exploring the possible interactions with higher
spins by constructing a higher spin supercurrent multiplet.

%It is known that any matter theory can be coupled to (super)gravity
%by following the well known minimal coupling procedure in order
%to introduce the gravitational (super)field. For that case the 
%calculation of the conserved (super)current is straightforward.
%One has to take the functional derivative of the interacting action
%with respect the gravitational (super)field and keep only the
%first order terms (see e.g.\cite{BK}). However, this approach is not
%applicable for higher spins because we do not know the fully interacting
%theory and the concept of minimal coupling is not relevant any more.
%The only alternative option we have is to follow Noether's method
%in order to construct directly the higher spin supercurrent multiplet
%of the theory.

It is known that any $\N=1$ supersymmetric matter theory can be consistently 
coupled to supergravity with the help of the gravitational superfield. For that 
case the calculation of the
conserved supercurrent is straightforward.  One has to take the functional 
derivative of the interacting action with respect the gravitational superfield  
(see e.g. \cite{BK}). However, this procedure is not applicable for higher spin 
theory  because we do not know the fully interacting theory at present.   The 
only alternative option we have is to follow
Noether's method in order to construct directly the higher spin supercurrent 
multiplet of the theory. However, in the case of coupling to supergravity we 
should be sure that the Noether procedure leads to the same supercurrent as the 
supergravity procedure.
%
%
%Unlike the $D=4,\,{\cal N}=1$ superspace formulation of supergravity,
%where there
%is a universal method to derive the supercurrent based on the use of the
%gravitational superfield (see e.g. \cite{BK}), in four dimensional
%higher spin field theory such a procedure is absent. Therefore the
%only possibility to construct the conserved higher spin supercurrent
%related to Noether procedure. Namely this procedure will be explored
%in this paper. However, using the superfield Noether procedure for
%constructing the supercurrent we should be sure that such a
%procedure leads to the known results of the supergravity method.

In this paper, we are following a Noether-type approach and we 
search for the higher spin supercurrent multiplet that generates 
the first order coupling of the interacting matter theory with 
the higher spin supermultiplets of type $(s+1,s+1/2)$. We find 
that interactions with higher spin supermultiplets beyond 
supergravity ($s\geq2$) are not possible for any $\K$ and $\W$ and 
thus extending the results of
the no-go Coleman-Mandula theorem\footnote{Examples of bypassing
the Coleman-Mandula theorem are discussed in \cite{Maldacena}}.
However, we find three exceptions to this rule and these are
(\emph{i}) a free massless chiral superfield, (\emph{ii}) a free
massive chiral superfield and (\emph{iii}) a free chiral superfield
with a linear superpotential. 
%These exceptions are the corresponding
%trivial $\mathcal{S}$- matrix exceptions of the Coleman-Mandula
%theorem. 
For the first two, the higher spin supercurrents and
supertraces have been constructed in
\cite{Superspace1,Superspace2,Superspace3,Superspace5}. We add to
this list the expressions for the supercurrent and supertrace for
the third theory. 

We also consider lower spin supercurrents. As mentioned previously,
unlike the higher spin case,
coupling of the theory under consideration with non-minimal 
supergravity can always be found for any $\K$ and $\W$.
Indeed, this follows from our analysis
and we get expression compatible with the results of \cite{Magro}.
On the other hand, interactions with the vector supermultiplet do
not always exist. We find the necessary and sufficient conditions
are the existence of a redefinition of the chiral superfield
$\Phi\to\varphi$ such that the chiral superpotential vanishes 
($\W=0$)
and the K\"{a}hler potential depends only in the product of
$\bar{\varphi}\varphi$ ($\K=\K(\bar{\varphi}\varphi)$). These
conditions can be
understood as the requirements for the presence of a global $U(1)$
symmetry which is usually associated with the vector supermultiplet.

The paper is organized as follows. In section two, we review Noether's method and
its application for the construction of first order in $g$ interaction vertices.
In addition, we review the description of free $4D,~\N=1$ higher spin
supermultiplets and the conservation equation of the supercurrent multiplet. In
section three, we focus on the vector supermultiplet and go through the
requirements in order to construct a conserved current out of the chiral theory.
In
section four, we repeat the procedure for supergravity and similarly in section
five for higher spin supermultiplets. In previous sections we had the chiral
superpotential $\W$ turned of. In section six, we turn it back on and consider
its contribution to the supercurrent multiplets. Finally, in section seven
we review and discuss our results.
%%%%%%%%%%%%%%%%%%%%%%%%%%%%%%%%%%%%%%%%
\section{Noether's method for supersymmetric nonlinear sigma model and higher spins}
\label{sec:review}
%%%%%%%%%%%%%%%%%%%%%%%%%%%%%%%%%%%%%%%%
The fundamental principles that govern higher spin interactions are still not
understood. Hence, the only guiding principle one has, is the physical
requirement of preserving the propagating degrees of freedom. This is manifested
through gauge invariance. Noether's method is the framework where one organizes
the invariance requirement order by order in a perturbative expansion around a
starting point $S_0$. In this approach the full action $S[
\phi,h]$ and transformation of fields $\phi,~h$ are expanded in a power series
of a coupling constant $g$:
\bea{l}
S[\phi,h]=S_0[\phi]+gS_1[\phi,h]+g^2S_2[\phi,h]+\dots~,\n\\
\delta \phi=\delta_0[\xi]+g\delta_1[\phi,\xi]+g^2\delta_2[\phi,\xi]+\dots~,\n\\
\delta h=\delta_0[\zeta]+g\delta_1[h,\zeta]+g^2\delta_2[h,\zeta]+\dots~.\n
\eea
The first order in $g$ interaction terms are given by $S_1$ and the requirement
of invariance for this order gives:
\bea{l} \n\label{Noether-1}
\frac{\delta S_0}{\delta\phi}\delta_1\phi+\frac{\delta S_1}{\delta
h}\delta_0h=0~.\n
\eea
In \cite{Superspace2} we demonstrated that for the case of a single chiral
superfield, most of the structure of $\d_1\Phi$ is fixed by the chiral requirement
($\Dd_{\ad}~\d_1\Phi=0$) and we explored the consequences of \eqref{Noether-1}
for the choice of $S_0$ corresponding to the free theory of a chiral superfield.
In this paper we want to explore if there are interaction terms $S_1$ that
correspond to a different starting action $S_0$.
% Because
%$\d_1\Phi$ is fixed, the only freedom we have is in the choice of starting
%action $S_0$.

In order to be as general as possible, we will consider as our starting point a
supersymmetric nonlinear sigma model described by an arbitrary K\"{a}hler
potential $\K(\Phi,\bar{\Phi})$ and a chiral superpotential $\W(\Phi)$
\bea{l}
S_{0}=\int d^8z~\K(\Phi,\bar{\Phi})+\int d^6z~\W(\Phi)
+\int d^6\bar{z}~\mathcal{\bar{W}}(\bar{\Phi})\n~
\eea
where the K\"{a}hler potential and the chiral superpotential are defined
modulo the relations
\bea{l}
\K(\Phi,\bar{\Phi})\sim\K(\Phi,\bar{\Phi})+\L(\Phi)+\bar{\L}(\bar{\Phi})~,\n\\
\W(\Phi)\sim\W(\Phi)+\textit{constant}~.\n\label{eqclW}
\eea
The on-shell equation of motion for this system is
\footnote{We follow the conventions of \emph{Superspace}\cite{GGRS}}
\bea{l}
\Dd^2\K_{\Phi}=\W_{\Phi}\n\label{eom}
\eea
and the invariance requirement \eqref{Noether-1} becomes:
\bea{l}
\int d^8z~\left\{\vphantom{\frac12}\K_{\Phi}~\d_1\Phi+\J\d_{0}h\right\}
+\int d^6z~\W_{\Phi}~\d_1\Phi
=0\n
\label{Noether-supercurrent}~.
\eea
The above expression is symbolic in the sense that $h$ corresponds to the set of
superfields that participate in the description of the $4D,~\N=1$ free,
massless, higher spin supermultiplets and also we have assumed that the
interaction terms can be written as higher spin gauge superfields times
corresponding elements of the conserved supercurrent multiplet $\J$.

The massless, higher spin irreducible representations of the super-Poincar\'{e}
group in four dimensions were first described in \cite{hss1}. Later, a superfield
formulation was introduced in \cite{hss2,hss3,hss4} and further developments can
be found in \cite{hss5,hss6,hss7}. A quick synopsis of the description of higher
spin supermultiplets is the following:
\begin{enumerate}
\item The integer superspin $Y=s$ ($s\geq1$) supermultiplets $(s+1/2 ,
s)$\footnote{On-shell they
describe the propagation of degrees of freedom with helicity $\pm (s+1/2)$ and
$\pm s$}
are described by a pair of superfields
$\Psi_{\a(s)\ad(s-1)}$\footnote{The notation $\a(k)$ is a shorthand for k
undotted symmetric indices $\a_1\a_2\dots\a_k$.
The same notation is used for the dotted indices }
 and $V_{\a(s-1)\ad(s-1)}$ with the following zero
order gauge transformations
\bea{l}\n\label{hstr1}
\d_0\Psi_{\a(s)\ad(s-1)}=-\D^2L_{\a(s)\ad(s-1)}+\tfrac{1}{(s-1)!}\Dd_{(\ad_{
s-1}}\Lambda_{\a(s)\ad(s-2))}~,\sn\\  \d_0 V_{\a(s-1)\ad(s-1)}=\D^{\a_s}L_{\a(
s)\ad(s-1)}+\Dd^{\ad_s}\bar{L}_{\a(s-1)\ad(s)}\sn~.
\eea
%%%%
\item The half-integer superspin $Y=s+1/2$ supermultiplets $(s+1 , s+1/2)$
have two descriptions. The first is called the transverse formulation ($s\geq1$)
and it uses the pair of superfields $H_{\a(s)
\ad(s)}$, $\chi_{\a(s)\ad(s-1)}$ with the following zero order gauge
transformations
\bea{l}\n\label{hstr2}
\d_0
H_{\a(s)\ad(s)}=\tfrac{1}{s!}\D_{(\a_s}\bar{L}_{\a(s-1))\ad(s)}-\tfrac{1}{s!}
\Dd_{(\ad_s}L_{\a(s)\ad(s-1))}\sn\label{hstr2H}~,\\
\d_0\chi_{\a(s)\ad(s-1)}=\Dd^2L_{\a(s)\ad(s-1)}+\D^{\a_{s+1}}\Lambda_{\a(
s+1)\ad(s-1)}~.\sn
\eea
The second one is the longitudinal formulation ($s\geq2$) and it includes the
superfields $H_{\a(s)\ad(s)}$, $\chi_{\a(s-1)\ad(
s-2)}$ with
\bea{l}\n\label{hstr3}
\d_0
H_{\a(s)\ad(s)}=\tfrac{1}{s!}\D_{(\a_s}\bar{L}_{\a(s-1))\ad(s)}-\tfrac{1}{s!}
\Dd_{(\ad_s}L_{\a(s)\ad(s-1))}\sn~,\vspace{1ex}\\
\d_0\chi_{\a(s-1)\ad(s-2)}=\Dd^{\ad_{s-1}}\D^{\a_s}L_{\a(s)\ad(s-1)}+\tfrac{
s-1}{s}\D^{\a_s}\Dd^{\ad_{s-1}}L_{\a(s)\ad(s-1)}\sn\\
\hspace{17ex}
+\tfrac{1}{(s-2)!}\Dd_{(\ad_{s-2}}J_{\a(s-1)\ad(s-3))}~.
\eea
\end{enumerate}
The $s=0$ case corresponds to the well known vector supermultiplet $(1,1/2)$
which is being described by a real scalar superfield $V$
with the gauge transformation $\d_0 V=\Dd^2 L+\D^2\bar{L}$.
The invariance of the action up to first order in $g$ as expressed in
\eqref{Noether-1} makes obvious that if we go on-shell ($\frac{\d
S_{0}}{\d\Phi}=0$ )
we get a conservation condition on the supercurrent multiplet $\J$ which is
controlled by the zeroth order gauge
transformation of the higher spin superfields. Using the expressions above,
we find the precise conservation conditions:
\begin{enumerate}
\item For integer superspin $Y=s$ ~~$(s+1/2 , s)$ we must have
\bea{l}\n
\D^2\Dd^{\ad_{s}}\J_{\a(s)\ad(s)}=\tfrac{1}{s!}\D_{(\a_{s}}\T_{\a(s-1))\ad(s-1)}
~~~,~~~\T_{\a(s-1)\ad(s-1)}=\bar{\T}_{\a(s-1)\ad(s-1)}~.
\eea
\item For transverse half-integer superspin $Y=s+1/2$ ~~$(s+1 , s+1/2)$
\bea{l}\n\label{ce1}
\Dd^{\ad_{s}}\J_{\a(s)\ad(s)}=\tfrac{1}{s!}\Dd^2\D_{(\a_{s}}\T_{\a(s-1))
\ad(s-1)}
~~~,~~~\J_{\a(s)\ad(s)}=\mathcal{\bar{J}}_{\a(s)\ad(s)}~.
\eea
\item For longitudinal half-integer superspin $Y=s+1/2$ ~~$(s+1 , s+1/2)$
\bea{l}\n\label{ce2}
\hspace{-5ex}\Dd^{\ad_{s}}\J_{\a(s)\ad(s)}=\tfrac{1}{s!}\D_{(\a_{s}}
\Dd^2\T_{\a(s-1))\ad(s-1)}
-\tfrac{s-1}{s!s!}\Dd_{(\ad_{s-1}}\D_{(\a_s}\Dd^{\bd}\T_{\a(s-1))\bd\ad(s-2))}
~,~\J_{\a(s)\ad(s)}=\mathcal{\bar{J}}_{\a(s)\ad(s)}~.
\eea
\end{enumerate}
The superfields $\J$ and $\T$ (with appropriate index structures) are the
higher spin supercurrent and higher spin supertrace respectively and together they
define the supercurrent multiplet which generate the first order interaction terms
with the higher spin gauge superfields:
\begin{enumerate}
\item For integer superspin $Y=s$ ~~$(s+1/2 , s)$
\bea{l}
S_1\sim\int
d^8z~\left\{\vphantom{\frac12}
\Psi^{\a(s)\ad(s-1)}\J_{\a(s)\ad(s-1)}
+\bar{\Psi}^{\a(s-1)\ad(s)}\mathcal{\bar{J}}_{\a(s-1)\ad(s)}
+V^{\a(s-1)\ad(s-1)}\T_{\a(s-1)\ad(s-1)}\right\}.~~~\n
\eea
\item For transverse half-integer superspin $Y=s+1/2$ ~~$(s+1 , s+1/2)$
\bea{l}
S_1\sim\int
d^8z\left\{\vphantom{\frac12}
H^{\a(s)\ad(s)}\J_{\a(s)\ad(s)}
+\chi^{\a(s)\ad(s-1)}\D_{\a_s}\T_{\a(s-1)\ad(s-1)}
+\bar{\chi}^{\a(s-1)\ad(s)}\Dd_{\ad_s}\mathcal{\bar{T}}_{\a(s-1)\ad(s-1)}\right\}.~~~\n
\label{Int}
\eea
\item For longitudinal half-integer superspin $Y=s+1/2$ ~~$(s+1 , s+1/2)$
\bea{l}
\hspace{-6ex}S_1\sim\hspace{-1ex}\int
d^8z\left\{\vphantom{\frac12}
H^{\a(s)\ad(s)}\J_{\a(s)\ad(s)}
+\chi^{\a(s-1)\ad(s-2)}\Dd^{\ad_{s-1}}\T_{\a(s-1)\ad(s-1)}
+\bar{\chi}^{\a(s-2)\ad(s-1)}\D^{\a_{s-1}}\mathcal{\bar{T}}_{\a(s-1)\ad(s-1)}\right\}.~~~~\n
\eea
\end{enumerate}
Furthermore, conservation equations \eqref{ce1} and \eqref{ce2} are not
independent. They are related via an improvement term
$X_{\a(s-1)\ad(s-1)}$. It is straight forward to show that if the superfields
$\J_{\a(s)\ad(s)}$,
$\T^{\perp}_{\a(s-1)\ad(s-1)}$ and $\T^{\parallel}_{\a(s-1)\ad(s-1)}$ satisfy
the following conservation equation
\bea{ll}\n\label{g.c.e}
\Dd^{\ad_{s}}\J_{\a(s)\ad(s)}=&~\tfrac{1}{s!}\Dd^2\D_{(\a_{s}}\T^{\perp}
_{\a(s-1))\ad(s-1)}\n\\
&+\tfrac{1}{s!}\D_{(\a_{s}}\Dd^2\T^{\parallel}_{\a(s-1))\ad(s-1)}
-\tfrac{s-1}{s!s!}\Dd_{(\ad_{s-1}}\D_{(\a_s}\Dd^{\bd}\T^{\parallel}_{\a(s-1))\bd
\ad(s-2))}
\eea
then the hatted superfields
\bea{l}\n\label{g.hat.multiplet}
\mathcal{\hat{J}}_{\a(s)\ad(s)}=\J_{\a(s)\ad(s)}+\tfrac{1}{s!s!}\Dd_{(\ad_{s}}
\D_{(\a_{s}}X_{\a(s-1))\ad(s-1))}
-\tfrac{1}{s!s!}\D_{(\a_{s}}\Dd_{(\ad_{s}}\bar{X}_{\a(s-1))\ad(s-1))}~,\sn\\
\hat{\T}^{\perp}_{\a(s-1)\ad(s-1)}=\T^{\perp}_{\a(s-1)\ad(s-1)}+\tfrac{s+1}{s}
X_{\a(s-1)\ad(s-1)}+\bar{X}_{\a(s-1)\ad(s-1)}~,\sn\\
\hat{\T}^{\parallel}_{\a(s-1)\ad(s-1)}=\T^{\parallel}_{\a(s-1)\ad(s-1)}+\bar{X}
_{\a(s-1)\ad(s-1)}~,\sn
\eea
satisfy exactly the same conservation equation.
%\bea{ll}\n\label{hat.g.c.e}
%\Dd^{\ad_{s}}\mathcal{\hat{J}}_{\a(s)\ad(s)}=&~\tfrac{1}{s!}\Dd^2\D_{(\a_{s}}
%\hat{\T}^{\perp}_{\a(s-1))\ad(s-1)}\n\\
%&+\tfrac{1}{s!}\D_{(\a_{s}}\Dd^2\hat{\T}^{\parallel}_{\a(s-1))\ad(s-1)}
%-\tfrac{s-1}{s!s!}\Dd_{(\ad_{s-1}}\D_{(\a_s}\Dd^{\bd}\hat{\T}^{\parallel}
%_{\a(s-1))\bd\ad(s-2))}
%\eea
So, there is a choice of $X_{\a(s-1)\ad(s-1)}$ that will convert \eqref{ce1}
[$\T^{\parallel}_{\a(s-1)\ad(s-1)}=0$] to \eqref{ce2}
[$\hat{\T}^{\perp}_{\a(s-1)\ad(s-1)}=0$] and another one to go from \eqref{ce2}
[$\T^{\perp}_{\a(s-1)\ad(s-1)}=0$] to \eqref{ce1}
[$\hat{\T}^{\parallel}_{\a(s-1)\ad(s-1)}=0$]. This is a manifestation of the
fact that the two formulations of half-integer superspin supermultiplets are
dual to each other.

Based on the results of \cite{Superspace2} we know that if 
$\d_1\Phi$ is linear in derivatives of $\Phi$\footnote{That is to be distinguished
from terms linear in derivatives of $\bar{\Phi}$.} then interactions with integer
superspin supermultiplets require more than one chiral supefields. Therefore, in
this paper we
will focus our efforts in constructing interactions with half integer superspin
supermultiplets ($s+1,s+1/2$) of the \eqref{Int} kind via higher spin 
supercurrent multiplets that satisfy conservation
equation \eqref{ce1}. However, in
order to get some intuition
and understand all the contributing factors we will not start with the arbitrary
spin case but  from $s=0$ (vector supermultiplet) to $s=1$ (supergravity) and
then to higher spin supermultiplets ($s\geq2$). Furthermore, in order to
avoid unnecessary complexity we will turn off $\W$ for the next three sections
and consider only the effects of $\K$. The contributions of $\W$ will be
examined in section six.
%%%%%%%%%%%%%%%%%%%%%%%%%%%%%%%%%%%%%%%%%%%%%%
\section{Coupling to vector supermultiplet}
\label{sec:vector}
%%%%%%%%%%%%%%%%%%%%%%%%%%%%%%%%%%%%%%%%%%%%%%
In this case, the conservation equation \eqref{ce1} gets simplified to
\bea{l}
\Dd^2\J=0\n\label{s=0:ce}
\eea
and the supercurrent multiplet has only one element, the real, scalar
supercurrent $\J$.
Due to (\ref{Noether-supercurrent}) and the structure of $\d_1\Phi$ as found
in \cite{Superspace2}, the supercurrent $\J$ must depend on $\Phi,\bar{\Phi}$
but crucially not in their derivatives.
Hence we should be able to express $\J$ as a power series
\bea{l}\n
\J=\sum_{p}\sum_{q}\Phi^{p}\bar{\Phi}^{q}A_{p,q}\n
\eea
where $A_{p,q}$ are a set of constants.
The conservation equation \eqref{s=0:ce} gives:
\bea{l}\n
\Dd^2\J=~\sum_{p}\Phi^{p}~\Dd^2\hspace{-0.5ex}\left[~\sum_{q}\bar{\Phi}^{q}
A_{p,q}~\right]=0~.
\eea
Furthermore, because it must hold on-shell ($\Dd^2\K_{\Phi}=0$), we must have
\bea{l}
\Dd^2\hspace{-0.5ex}\left.\left[~\sum_{q}\bar{\Phi}^{q}A_{p,q}~\right]
\right|_{\Dd^2\K_{\Phi}=0}=0
~\Rightarrow~\sum_{q}\bar{\Phi}^{q}A_{p,q}=f_{p}(\Phi)\K_{\Phi}\n
\eea
where $f_{p}(\Phi)$ is a function of $\Phi$. Hence, we conclude that $\J$ must
be of the form
\bea{l}\n
\J=\sum_{p}\Phi^{p}f_{p}(\Phi)~\K_{\Phi} = F(\Phi)\K_{\Phi}
\eea
where $F(\Phi)=\sum\limits_{p}\Phi^p f_{p}(\Phi)$.
However, $\J$ by definition has to be real therefore we must have
\bea{l}
F(\Phi)\K_{\Phi}=\bar{F}(\bar{\Phi})\bar{\K}_{\bar{\Phi}}~.\n
\eea
For $F(\Phi)\neq0$ which is the non-trivial case we are being interested, we
can define a new chiral superfield $\varphi$ as follows:
\bea{l}
\varphi=exp{\left[\int d\Phi ~F^{-1}(\Phi)\right]}~.\n
\eea
For this new variable the on-shell equation of motion has the same form as
before ($\Dd^2\K_{\varphi}=0$), the supercurrent $\J$ takes the form
\bea{l}
\J=\varphi\K_{\varphi}\n\label{s=0:J}
\eea
and the reality condition takes the simpler expression
\bea{l}
\varphi\K_{\varphi}=\bar{\varphi}\K_{\bar{\varphi}}~.\n
\eea
This can be satisfied only if $\K$ is a function of the product
$\bar{\varphi}\varphi$
\footnote{Equivalently, one can define another chiral superfield
$\phi=ln(\varphi)=\int d\Phi ~F^{-1}(\Phi)$ under which the K\"{a}hler potential
is a function of the sum $\phi+\bar{\phi}$, ~$\K=\K(\phi+\bar{\phi})$ and the
supercurrent is $\J=\K_{\phi}$. A detailed discussion of this can be found
in \cite{GGRS} where the connection between the action of a real linear $G=\phi+\bar{\phi}$
and the action of a chiral $\varphi$ in presented.}, $\K=\K(\bar{\varphi}\varphi)$.
%\bea{l}
%\K=\K(\bar{\varphi}\varphi)\n\label{s=0:K}~.
%\eea
This constraint in $\K$ can be understood as the demand for the K\"{a}hler
potential to have a global $U(1)$ symmetry expressed by the phase shift
of $\varphi$
\footnote{For the variable $\phi$ the global $U(1)$ is realized as a shift
symmetry $\phi \to \phi+i\lambda$.}
\bea{l}
\varphi \to e^{i\lambda}\varphi\n
\eea
which can be gauged in order to generate interactions with the vector
supermultiplet.
%%%%%%%%%%%%%%%%%%%%%%%%%%%%%%%%%%%%%%%%%%%%%%%%%%%
\section{Coupling to (non-minimal) supergravity}
\label{sec:sugra}
%%%%%%%%%%%%%%%%%%%%%%%%%%%%%%%%%%%%%%%%%%%%%%%%%%%
Next, we attempt the construction of the supercurrent multiplet that generates
interactions with non-minimal supergravity [transverse formulation of
supermultiplet $(2,3/2)$] which satisfies the conservation equation
\bea{l}
\Dd^{\ad}\J_{\a\ad}=\Dd^2\D_{\a}\T~.\n\label{s=1:ce}
\eea
For the supercurrent and supertrace we consider the following ansatz:
\bea{lll}
\J_{\a\ad}=
&~~\delta~\pa_{\a\ad}\Phi~\K_{\Phi}
~~&+d~\pa_{\a\ad}\bar{\Phi}~\K_{\bar{\Phi}}\n\label{s=1:J}\\
&+\a~\D_{\a}\Phi~\Dd_{\ad}\K_{\Phi}
~~&-a~\Dd_{\ad}\bar{\Phi}~\D_{\a}\K_{\bar{\Phi}}\\
&+\b~\Phi~\D_{\a}\Dd_{\ad}\K_{\Phi}
~~&-b~\bar{\Phi}~\Dd_{\ad}\D_{\a}\K_{\bar{\Phi}}\\
&+\g~\Phi~\Dd_{\ad}\D_{\a}\K_{\Phi}
~~&-c~\bar{\Phi}~\D_{\a}\Dd_{\ad}\K_{\bar{\Phi}}\vspace{1ex}\\
\T=&e~\K+\kappa~\Phi~\K_{\Phi}&+h~\bar{\Phi}~\K_{\bar{\Phi}}~.\n\label{s=1:T}
\eea
%*************************
\subsection{Improvement terms}
%*************************
The definition of the supercurrent multiplet $(\J_{\a\ad}, \T)$ via conservation
equation \eqref{s=1:ce} is not unique.
There are improvement terms that one should consider. In this case, there is
an arbitrary superfield $U_{\a}$ such that
the supercurrent multiplet $(\mathcal{\hat{J}}_{\a\ad},\mathcal{\hat{T}})$
defined by:
\bea{l}\n\label{new.J.T}
\mathcal{\hat{J}}_{\a\ad}=\J_{\a\ad}+\D_{\a}\Dd^2\bar{U}_{\ad}-\Dd_{\ad}
\D^2U_{\a}~,\sn\\
\mathcal{\hat{T}}=\T+2\D^{\a}U_{\a}+\Dd^{\ad}\bar{U}_{\ad}\sn
\eea
satisfies the same conservation equation \eqref{s=1:ce}. This can also be
extracted from
(\ref{g.c.e},\ref{g.hat.multiplet}) and the demand that
the hat supercurrent and supertrace stay in the transverse formulation.
If we select
\bea{l}
U_{\a}=r~\D_{\a}\bar{\Lambda}~\K_{\bar{\Phi}}\n\label{improve}
\eea
where $\Lambda$ is the prepotential of chiral superfield $\Phi=\Dd^2\Lambda$,
then the $r$ parameter can be used to eliminate $b$ in \eqref{s=1:J}.
Hence the ansatz for $\J_{\a\ad}$ becomes
\bea{lll}
\J_{\a\ad}=
&~~\delta~\pa_{\a\ad}\Phi~\K_{\Phi}
~~&+d~\pa_{\a\ad}\bar{\Phi}~\K_{\bar{\Phi}}\n\label{s=1:Jb}\\
&+\a~\D_{\a}\Phi~\Dd_{\ad}\K_{\Phi}
~~&-a~\Dd_{\ad}\bar{\Phi}~\D_{\a}\K_{\bar{\Phi}}\\
&+\b~\Phi~\D_{\a}\Dd_{\ad}\K_{\Phi}
~~& \\
&+\g~\Phi~\Dd_{\ad}\D_{\a}\K_{\Phi}
~~&-c~\bar{\Phi}~\D_{\a}\Dd_{\ad}\K_{\bar{\Phi}}~.
\eea
%*******************************
\subsection{Conservation equation}
%*******************************
Now we use \eqref{s=1:Jb} and \eqref{s=1:T} to determine the consequences of
conservation equation \eqref{s=1:ce}.
The result is:
\bea{l}\n\label{ce.const.s=1}
0=\Dd^{\ad}\J_{\a\ad}-\Dd^2\D_{\a}\T=
\left(i\d+\a+\kappa+e\right)~\Dd^{\ad}\D_{\a}\Phi~\Dd_{\ad}\K_{\Phi}~~+
\left(2\g-\b-\kappa\right)~\Phi~\Dd^2\D_{\a}\K_{\Phi}\\
\hspace{23ex}+id~\D_{\a}\Dd^{2}\bar{\Phi}~\K_{\bar{\Phi}}~~+id~\D_{\a}\Dd^{\ad}
\bar{\Phi}~\Dd_{\ad}\K_{\bar{\Phi}}\\
\hspace{23ex}-\left(2a+h\right)~\Dd^2\bar{\Phi}~\D_{\a}\K_{\bar{\Phi}}~~-\left(a
+h\right)~\Dd^{\ad}\bar{\Phi}~\Dd_{\ad}\D_{\a}\K_{\bar{\Phi}}\\
\hspace{23ex}-c~\Dd^{\ad}\bar{\Phi}~\D_{\a}\Dd_{\ad}\K_{\bar{\Phi}}~~-\left(h-c
\right)~\bar{\Phi}~\Dd^2\D_{\a}\K_{\bar{\Phi}}\\
\hspace{23ex}+c~\bar{\Phi}~\D_{\a}\Dd^{2}\K_{\bar{\Phi}}~.
\eea
Assuming that the K\"{a}hler potential $\K$ is arbitrary, meaning it does not
have special properties that relate some of the terms above with each other
and they are independent,
we conclude that the coefficient of each term must
vanish:
\bea{l}
e=-i\d-\a-2\g+\b~,\\
\kappa=2\g-\b~,\\
a=0~,\n\\
c=0~,\\
d=0~,\\
h=0
\eea
and the expressions for the conserved $\J_{\a\ad}$ and $\T$ are:
\bea{l}
\J_{\a\ad}=\delta~\pa_{\a\ad}\Phi~\K_{\Phi}+\a~\D_{\a}\Phi~\Dd_{\ad}\K_{\Phi}+
\b~\Phi~\D_{\a}\Dd_{\ad}\K_{\Phi}+\g~\Phi~\Dd_{\ad}\D_{\a}\K_{\Phi}~,\n
\label{s=1:cJ}\\
\T=(-i\d-\a-2\g+\b)~\K+(2\g-\b)~\Phi~\K_{\Phi}~.\n
\eea
%******************************
\subsection{Reality Condition}
%******************************
The last condition we must impose is the reality of $\J_{\a\ad}$. For this it will
be useful to take into account the following expressions
\bea{l}
\Dd_{\ad}\K_{\Phi}=\Dd_{\ad}\bar{\Phi}~\K_{\Phi\bar{\Phi}}~,\\
\D_{\a}\Dd_{\ad}\K_{\Phi}=i\pa_{\a\ad}\bar{\Phi}~\K_{\Phi\bar{\Phi}}+\D_{\a}
\Phi~\Dd_{\ad}\bar{\Phi}~\K_{\Phi\Phi\bar{\Phi}}~,\n\label{expK}\\
\Dd_{\ad}\D_{\a}\K_{\Phi}=i\pa_{\a\ad}\Phi~\K_{\Phi\Phi}-\D_{\a}\Phi~\Dd_{\ad}
\bar{\Phi}~\K_{\Phi\Phi\bar{\Phi}}
\eea
which can be used to re-write \eqref{s=1:cJ} in the following manner
\bea{l}
\J_{\a\ad}=~~\pa_{\a\ad}\Phi~\left[\d~\K_{\Phi}+i\g~\Phi~\K_{\Phi\Phi}\right]\n\label{dc1}\\
\hspace{6.5ex}+\pa_{\a\ad}\bar{\Phi}~\left[i\b~\Phi~\K_{\Phi\bar{\Phi}}\right]\\
\hspace{6.5ex}+\D_{\a}\Phi~\Dd_{\ad}\bar{\Phi}~\left[\a~\K_{\Phi\bar{\Phi}}+(\b-
\g)~\Phi~\K_{\Phi\Phi\bar{\Phi}}\right]~.
\eea
The reality of $\J_{\a\ad}$ as expressed above demands:
\bea{l}
1.)~~~~~~\d~\K_{\Phi}+i\g~\Phi~\K_{\Phi\Phi}+i\b^*~\bar{\Phi}~\K_{\Phi
\bar{\Phi}}=0~,\n\vspace{1ex}\\
2.)~~~~~\left(\a^*-\a\right)~\K_{\Phi\bar{\Phi}}-\left(\b-\g\right)~\Phi~
\K_{\Phi\Phi\bar{\Phi}}+\left(\b-\g\right)^*~\bar{\Phi}~\K_{\Phi\bar{\Phi}
\bar{\Phi}}=0~.\n
\eea
For an arbitrary K\"{a}hler
potential, the above constraints can be satisfied only if
\bea{l}
\a=\a^*~,~\b=\g=\d=0~.\n
\eea
Therefore, we conclude that for an arbitrary K\"{a}hler potential $\K$ there
exist a supercurrent multiplet
\bea{l}
\J_{\a\ad}=\D_{\a}\Phi~\Dd_{\ad}\K_{\Phi}~,\n\label{J1}\\
\T=-~\K\n\label{T1}~.
\eea
Conceptually, this result was to be expected because we know
that any theory can be coupled to supergravity. Expressions,
\eqref{J1} and \eqref{T1} give the supercurrent multiplet that
generates the linearized interaction (first order in $g$)
between a supersymmetric nonlinear model of a single chiral
superfield described by K\"{a}hler potential $\K$ and
non-minimal supergravity. Additionally, we observe that for
arbitrary $\K$ the supertrace $\T$ is not zero and for that
case the supercurrent multiplet defined by $\{\J_{\a\ad},\T\}$
is \emph{canonical} \cite{Magro,Superspace2}. However, it is
straight forward to see that if the K\"{a}hler potential has
the property
\bea{l}
\K\sim \Phi\K_{\Phi}\n\label{cond.minimal}
\eea
then there is an improvement term of type \eqref{improve} such that the improved
supertrace $\mathcal{\hat{T}}$ \eqref{new.J.T} will be zero
\bea{l}
\mathcal{\hat{T}}=0
\eea
and the new supercurrent multiplet $\{\mathcal{\hat{J}}_{\a\ad},0\}$
is a \emph{minimal} one. This is possible only if the
K\"{a}hler potential is a function of the product $\bar{\Phi}\Phi$,~
$\K=\K(\bar{\Phi}\Phi)$, which includes the free theory.
Furthermore, by converting \eqref{J1} and \eqref{T1} via
\eqref{g.hat.multiplet} to the longitudinal formulation of
supergravity (minimal supergravity) we recover the results of \cite{Magro}.
%%%%%%%%%%%%%%%%%%%%%%%%%%%%%%%%%%%%%%%%%%%%%%%%%%%%%%
\section{Coupling to higher spin supermultiplets}
\label{sec:hs}
%%%%%%%%%%%%%%%%%%%%%%%%%%%%%%%%%%%%%%%%%%%%%%%%%%%%%%
Based on the two previous lower spin examples we build confidence on the
workings of our method and it is time to generalize it to higher spin
supermultiplet $(s+1,s+1/2)$ with $s\geq2$. For this case the conservation equation
we must satisfy is \eqref{ce1} and we write the following ansatz for the supercurrent
and the supertrace
\footnote{For simplicity, we omit to write explicitly the free indices
and their symmetrization when necessary. Also the symbol $\pa^{(p)}$
is an abbreviation for a string of $p$ spacetime derivatives.}:
\bea{lll}\n\label{bJs}
\J_{\a(s)\ad(s)}=
&~~\delta~\pa^{(s)}\Phi~\K_{\Phi}
~~&+d~\pa^{(s)}\bar{\Phi}~\K_{\bar{\Phi}}\\
&+\sum_{p=0}^{s-1} \a_{p}~\pa^{(p)}\D\Phi~\pa^{(s-p-1)}\Dd\K_{\Phi}
~~&-\sum_{p=0}^{s-1}
a_{p}~\pa^{(p)}\Dd\bar{\Phi}~\pa^{(s-p-1)}\D\K_{\bar{\Phi}}\\
&+\sum_{p=0}^{s-1} \b_{p}~\pa^{(p)}\Phi~\pa^{(s-p-1)}\D\Dd\K_{\Phi}
~~&-\sum_{p=0}^{s-1}
b_{p}~\pa^{(p)}\bar{\Phi}~\pa^{(s-p-1)}\Dd\D\K_{\bar{\Phi}}\\
&+\sum_{p=0}^{s-1} \g_{p}~\pa^{(p)}\Phi~\pa^{(s-p-1)}\Dd\D\K_{\Phi}
~~&-\sum_{p=0}^{s-1}
c_{p}~\pa^{(p)}\bar{\Phi}~\pa^{(s-p-1)}\D\Dd\K_{\bar{\Phi}}~~,
\eea
\bea{lll}\n\label{bTs}
\T_{\a(s-1)\ad(s-1)}=
&~~e~\pa^{(s-1)}\K~+~\kappa~\pa^{(s-1)}\Phi~\K_{\Phi}
~~&+h~\pa^{(s-1)}\bar{\Phi}~\K_{\bar{\Phi}}\\
&+\sum_{p=0}^{s-2} \ell_{p}~\pa^{(p)}\D\Phi~\pa^{(s-p-2)}\Dd\K_{\Phi}
~~&+\sum_{p=0}^{s-2}
f_{p}~\pa^{(p)}\Dd\bar{\Phi}~\pa^{(s-p-2)}\D\K_{\bar{\Phi}}\\
&+\sum_{p=0}^{s-2} \zeta_{p}~\pa^{(p)}\Phi~\pa^{(s-p-2)}\Dd\D\K_{\Phi}
~~&+\sum_{p=0}^{s-2}
g_{p}~\pa^{(p)}\bar{\Phi}~\pa^{(s-p-2)}\Dd\D\K_{\bar{\Phi}}\\
&+\sum_{p=0}^{s-2} \xi_{p}~\pa^{(p)}\Phi~\pa^{(s-p-2)}\D\Dd\K_{\Phi}
~~&+\sum_{p=0}^{s-2}
s_{p}~\pa^{(p)}\bar{\Phi}~\pa^{(s-p-2)}\D\Dd\K_{\bar{\Phi}}~~.
\eea
However, because $s\geq2$, the first term of $\T_{\a(s-1)\ad(s-1)}$
can be expanded in the following manner:
\bea{l}
\pa^{(s-1)}\K=\pa^{(s-2)}\left(~\pa\Phi~\K_{\Phi}+\pa\bar{\Phi}~\K_{\bar{\Phi}}
~\right)\\
\hspace{8.5ex}=\pa^{(s-1)}\Phi~\K_{\Phi}+\pa^{(s-1)}\bar{\Phi}~\K_{\bar{\Phi}}\n
\label{expand.dK}\\
\hspace{11ex}-i\sum_{p=1}^{s-2}\tbinom{s-2}{p-1}~\pa^{(p)}\Phi~\pa^{(s-p-2)}
\left(\D\Dd+\Dd\D\right)\K_{\Phi}~
-i\sum_{p=1}^{s-2}\tbinom{s-2}{p-1}~\pa^{(p)}\bar{\Phi}~\pa^{(s-p-2)}\left(\D
\Dd+\Dd\D\right)\K_{\bar{\Phi}}~.
\eea
Hence, this term is not independent any more, as in the supergravity case, and
all it does is to redefine the $\kappa,~h,~\zeta_{p},~g_{p},~ \xi_{p}$ and
$s_{p}$ coefficients. Therefore, we can ignore it ($e=0$). Moreover, both
$\J_{\a(s)\ad(s)}$
and $\T_{\a(s-1)\ad(s-1)}$ are not uniquely defined but up to an equivalence class.
This is due to the presence of terms that identically vanish in both left
and right hand sides of \eqref{ce1} due to the $\D$ algebra. For $\T_{\a(s-1)\ad(s-1)}$ this equivalence relation is
the following:
\bea{l}
\T_{\a(s-1)\ad(s-1)}\sim\T_{\a(s-1)\ad(s-1)}+\tfrac{1}{(s-1)!}\D_{(\a_{s-1}}
P^{(1)}_{\a(s-2))\ad(s-1)}+\D^{2}P^{(2)}_{\a(s-1)\ad(s-1)}
+\Dd^2P^{(3)}_{\a(s-1)\ad(s-1)}\n\label{eq.clas.T}
\eea
for arbitrary superfields $P^{(1)}_{\a(s-2)\ad(s-1)},~P^{(2)}_{\a(s-1)\ad(s-1)},
~P^{(3)}_{\a(s-1)\ad(s-1)}$. This means that we can immediately ignore the $
\ell_{p}$ terms because they can be converted to $\xi_{p}$ terms. Also, the
$f_{p}$ terms can be ignored because they can be converted to $g_{p}$ and $s_{p}
$ terms and additionally all the $s_{p}$ terms can be disregarded.
% Similarly for
%$\J_{\a(s)\ad(s)}$ we have:
%\bea{l}
%\J_{\a(s)\ad(s)}\sim\J_{\a(s)\ad(s)}+\Dd^{\ad_{s+1}}\Xi^{(1)}_{\a(s)\ad(s+1)}
%+\Dd^2\Xi^{(2)}_{\a(s)\ad(s)}\n\label{eq.clas.J}
%\eea
Therefore, the ansatz for the higher spin supertrace takes the form:
%\bea{lll}\n\label{bJs2}
%\J_{\a(s)\ad(s)}=
%&~~\delta~\pa^{(s)}\Phi~\K_{\Phi}
%~~&+d~\pa^{(s)}\bar{\Phi}~\K_{\bar{\Phi}}\\
%&+\sum_{p=0}^{s-1} \a_{p}~\pa^{(p)}\D\Phi~\pa^{(s-p-1)}\Dd\K_{\Phi}
%~~&-\sum_{p=0}^{s-1}
%a_{p}~\pa^{(p)}\Dd\bar{\Phi}~\pa^{(s-p-1)}\D\K_{\bar{\Phi}}\\
%&+\sum_{p=0}^{s-1} \b_{p}~\pa^{(p)}\Phi~\pa^{(s-p-1)}\D\Dd\K_{\Phi}
%~~&-\sum_{p=0}^{s-1}
%b_{p}~\pa^{(p)}\bar{\Phi}~\pa^{(s-p-1)}\Dd\D\K_{\bar{\Phi}}\\
%&+\sum_{p=0}^{s-1} \g_{p}~\pa^{(p)}\Phi~\pa^{(s-p-1)}\Dd\D\K_{\Phi}
%~~&-\sum_{p=0}^{s-1}
%c_{p}~\pa^{(p)}\bar{\Phi}~\pa^{(s-p-1)}\D\Dd\K_{\bar{\Phi}}
%\eea
\bea{lll}\n\label{bTs2}
\T_{\a(s-1)\ad(s-1)}=
&~~\kappa~\pa^{(s-1)}\Phi~\K_{\Phi}
~~&+h~\pa^{(s-1)}\bar{\Phi}~\K_{\bar{\Phi}}\\
&+\sum_{p=0}^{s-2} \zeta_{p}~\pa^{(p)}\Phi~\pa^{(s-p-2)}\Dd\D\K_{\Phi}
~~&+\sum_{p=0}^{s-2}
g_{p}~\pa^{(p)}\bar{\Phi}~\pa^{(s-p-2)}\Dd\D\K_{\bar{\Phi}}\\
&+\sum_{p=0}^{s-2} \xi_{p}~\pa^{(p)}\Phi~\pa^{(s-p-2)}\D\Dd\K_{\Phi}
~~&~~.
\eea
%*****************************************
\subsection{Improvement terms}
%*****************************************
Now we consider various improvement terms that will further reduce the unknown parameters in the above
expressions. The arguments that led to \eqref{new.J.T} also hold for $s\geq2$ as
well. Thus, the
improvement terms we
have are parametrized by an unconstrained superfield $U_{\a(s)\ad(s-1)}$ and
define the improved supercurrent and supertrace as follows:
\bea{l}\n
\mathcal{\hat{J}}_{\a(s)\ad(s)}=\J_{\a(s)\ad(s)}+\tfrac{1}{s!}~\D_{(\a_s}\Dd^2\bar{U}_{\a(s-1))\ad(s)}
-\tfrac{1}{s!}~\Dd_{(\ad_{s}}\D^2 U_{\a(s)\ad(s-1))}~,\sn\\
\mathcal{\hat{T}}_{\a(s-1)\ad(s-1)}=\T_{\a(s-1)\ad(s-1)}+\tfrac{s+1}{s}~\D^{\a_s}U_{\a(s)\ad(s-1)}
+~\Dd^{\ad_s}\bar{U}_{\a(s-1)\ad(s)}~.\sn
\eea
One can show that if we select $U_{\a(s)\ad(s-1)}$ in the following way
\bea{l}
U_{\a(s)\ad(s-1)}=r~\pa^{(s-1)}\D\bar{\Lambda}~\K_{\bar{\Phi}}\n\\
\hspace{12ex}+\sum_{p=0}^{s-2}~\r_{p}~\pa^{(p)}\D\bar{\Lambda}~\pa^{(s-p-2)}\Dd\D\K_{\bar{\Phi}}\\
\hspace{12ex}+\sum_{p=0}^{s-2}~\s_{p}~\pa^{(p)}\D\bar{\Lambda}~\pa^{(s-p-2)}\D\Dd\K_{\bar{\Phi}}~,
\eea
for some parameters $r,\r_{p},\s_{p}$, then
\footnote{Notice that the $\s_{p}$ terms do not participate in the result. That is because they describe
the freedom in the definition of $U_{\a(s)\ad(s-1)}$.}
\bea{lll}
\mathcal{\hat{J}}_{\a(s)\ad(s)}-\J_{\a(s)\ad(s)}=&~r~\pa^{(s-1)}\bar{\Phi}~\Dd\D\K_{\bar{\Phi}}~&~
-r^*~\pa^{(s-1)}\Phi~\D\Dd\K_{\Phi}\\
&+\sum_{p=0}^{s-2}~i\r_{p}~\pa^{(p)}\bar{\Phi}~\pa^{(s-p-1)}\Dd\D\K_{\bar{\Phi}}~&~
+\sum_{p=0}^{s-2}~i\r^{*}_{p}~\pa^{(p)}\Phi~\pa^{(s-p-1)}\D\Dd\K_{\Phi}\n\\
%%%%
&+r~\pa^{(s-1)}\Dd\bar{\Phi}~\D\K_{\bar{\Phi}}~&~
-r^*~\pa^{(s-1)}\D\Phi~\Dd\K_{\Phi}\\
&+\sum_{p=0}^{s-2}~i\r_{p}~\pa^{(p)}\Dd\bar{\Phi}~\pa^{(s-p-1)}\D\K_{\bar{\Phi}}~&~
+\sum_{p=0}^{s-2}~i\r^{*}_{p}~\pa^{(p)}\D\Phi~\pa^{(s-p-1)}\Dd\K_{\Phi}~.
\eea
As a result, we can select parameters $r$ and $\r_{p}$ in order to eliminate the $b_{p}$ terms in
\eqref{bJs}. Thus we get
\bea{lll}\n\label{bJs2}
\J_{\a(s)\ad(s)}=
&~~\delta~\pa^{(s)}\Phi~\K_{\Phi}
~~&+d~\pa^{(s)}\bar{\Phi}~\K_{\bar{\Phi}}\\
&+\sum_{p=0}^{s-1} \a_{p}~\pa^{(p)}\D\Phi~\pa^{(s-p-1)}\Dd\K_{\Phi}
~~&-\sum_{p=0}^{s-1}
a_{p}~\pa^{(p)}\Dd\bar{\Phi}~\pa^{(s-p-1)}\D\K_{\bar{\Phi}}\\
&+\sum_{p=0}^{s-1} \b_{p}~\pa^{(p)}\Phi~\pa^{(s-p-1)}\D\Dd\K_{\Phi}
~~& \\
&+\sum_{p=0}^{s-1} \g_{p}~\pa^{(p)}\Phi~\pa^{(s-p-1)}\Dd\D\K_{\Phi}
~~&-\sum_{p=0}^{s-1}
c_{p}~\pa^{(p)}\bar{\Phi}~\pa^{(s-p-1)}\D\Dd\K_{\bar{\Phi}}~.
\eea
%**********************************************************
\subsection{Additional freedom}
%**********************************************************
For $s\geq2$ there is some additional freedom in defining the higher spin supercurrent and supertrace.
Let us consider the following quantity:
\bea{l}
Z_{\a(s-1)\ad(s-1)}=\pa^{(s-2)}\left(\vphantom{\frac12}~\pa\bar{\Phi}\K_{\bar{\Phi}}
+i~\Phi\D\Dd\K_{\Phi}~\right)~.\n\label{extra.freedom}
\eea
It is straight forward to prove that
\bea{l}
\D_{(\a_{s}}Z_{\a(s-1))\ad(s-1)}=i\pa^{(s-2)}\left(\vphantom{\frac12}~\D\Phi~\D\Phi
~\Dd\bar{\Phi}~\K_{\Phi\Phi\bar{\Phi}}~\right)=0~.\n
\eea
It vanishes due to the symmetrization of the two $\D\Phi$ terms. Hence, we can enhance
the equivalence class in the definition of $\J_{\a(s)\ad(s)}$ and $\T_{\a(s-1)\ad(s-1)}$ by adding the
following terms
\bea{l}
\T_{\a(s-1)\ad(s-1)}\sim\T_{\a(s-1)\ad(s-1)}+c_1~Z_{\a(s-1)\ad(s-1)}~,\n\label{eq.clas.T.2}\\
\J_{\a(s)\ad(s)}\sim\J_{\a(s)\ad(s)}+c_2~\Dd_{(\ad_{s}}\D_{(\a_{s}}Z_{\a(s-1))\ad(s-1)}
+c_3~\D_{(\a_{s}}\Dd_{(\ad_{s}}\bar{Z}_{\a(s-1))\ad(s-1)}~.\n\label{eq.clas.J.2}
\eea
Because the $c_2$ and $c_3$ terms are identically zero they do not change $\J_{\a(s)\ad(s)}$,
whereas the $c_1$ term is not zero but does not contribute in the conservation equation.
If we expand these terms in a manner similar to \eqref{expand.dK} we get that the $h$ term
in \eqref{bTs2} can be set to zero by $c_1$, the $\a_{s-1}$ term in \eqref{bJs2} can be set
to zero by $c_3$ and similarly the $a_{s-1}$ term by $c_2$. Hence, we should consider the
following expressions for the higher spin supercurrent multiplet:
\bea{lll}\n\label{bJs3}
\J_{\a(s)\ad(s)}=
&~~\delta~\pa^{(s)}\Phi~\K_{\Phi}
~~&+d~\pa^{(s)}\bar{\Phi}~\K_{\bar{\Phi}}\\
&+\sum_{p=0}^{s-2} \a_{p}~\pa^{(p)}\D\Phi~\pa^{(s-p-1)}\Dd\K_{\Phi}
~~&-\sum_{p=0}^{s-2}
a_{p}~\pa^{(p)}\Dd\bar{\Phi}~\pa^{(s-p-1)}\D\K_{\bar{\Phi}}\\
&+\sum_{p=0}^{s-1} \b_{p}~\pa^{(p)}\Phi~\pa^{(s-p-1)}\D\Dd\K_{\Phi}
~~& \\
&+\sum_{p=0}^{s-1} \g_{p}~\pa^{(p)}\Phi~\pa^{(s-p-1)}\Dd\D\K_{\Phi}
~~&-\sum_{p=0}^{s-1}
c_{p}~\pa^{(p)}\bar{\Phi}~\pa^{(s-p-1)}\D\Dd\K_{\bar{\Phi}}~~,
\eea
\bea{lll}\n\label{bTs3}
\T_{\a(s-1)\ad(s-1)}=
&~~\kappa~\pa^{(s-1)}\Phi~\K_{\Phi}
~~& \\
&+\sum_{p=0}^{s-2} \zeta_{p}~\pa^{(p)}\Phi~\pa^{(s-p-2)}\Dd\D\K_{\Phi}
~~&+\sum_{p=0}^{s-2}
g_{p}~\pa^{(p)}\bar{\Phi}~\pa^{(s-p-2)}\Dd\D\K_{\bar{\Phi}}\\
&+\sum_{p=0}^{s-2} \xi_{p}~\pa^{(p)}\Phi~\pa^{(s-p-2)}\D\Dd\K_{\Phi}
~~&~~.
\eea
%**********************************************
\subsection{Conservation equation}
%**********************************************
The above streamlined expressions do not include any trivial parts for the
higher spin supercurrent and supertrace and are the ones we should use
with the conservation equation.
After a lengthy calculation and assuming once again that the
K\"{a}hler potential is arbitrary, we obtain the following system
of conditions:
\bea{l}
1.)~~~~~i\d+\kappa=0\vspace{0.7ex}~~,\\
2.)~~~~~\a_{p}~\left[\tfrac{s-p}{p}\right]-\b_{p+1}~\left[\tfrac{p+1}{s}\right]+i\xi_{p}=0~,
~~~p=0,1,\dots ,~s-2~~,\vspace{0.7ex}\\
3.)~~~~~\a_{s-2}~\left[\tfrac{1}{s}\right]-\b_{s-1}+\g_{s-1}~\left[\tfrac{s+1}{s}\right]-\kappa
+i\xi_{s-2}-i\zeta_{s-2}=0~~,\vspace{0.7ex}\\
4.)~~~~~\a_{p-1}~\left[\tfrac{s-p}{s}\right]-\b_{p}~\left[\tfrac{p+1}{s}\right]
+\g_{p}~\left[\tfrac{s+1}{s}\right]+i\xi_{p-1}-i\zeta_{p-1}-i\zeta_{p}=0~,
~~~p=1,\dots ,~s-2~~,\vspace{0.7ex}\\
5.)~~~~~\b_{0}~\left[\tfrac{1}{s}\right]-\g_{0}~\left[\tfrac{s+1}{s}\right]+i\zeta_{0}=0~~,\vspace{0.7ex}\\
6.)~~~~~a_{p}~\left[\tfrac{s+1}{s}\right]+ig_{p}=0~,~~~p=0,\dots ,~s-2~~,\vspace{0.7ex}\n\label{conserv.cond}\\
7.)~~~~~a_{p}~\left[\tfrac{p+1}{s}\right]+ig_{p}=0~,~~~p=0,\dots ,~s-2~~,\vspace{0.7ex}\\
8.)~~~~~c_{s-1}=0~~,\vspace{0.7ex}\\
9.)~~~~~c_{p}~\left[\tfrac{p+1}{s}\right]-ig_{p}=0~,~~~p=0,\dots ,~s-2~~,\vspace{0.7ex}\\
10.)~~~~d=0~~,\vspace{0.7ex}\\
11.)~~~~c_{p}=0~,~~~p=0,\dots ,~s-1~~,\vspace{0.7ex}\\
12.)~~~~a_{p}+c_{p}=0~,~~~p=0,\dots ,~s-1~~.
\eea
These uniquely fix all parameters except $\a_{p},~\b_{p}$ and $\g_{p}$. Specifically we get
the following solutions:
\bea{l}
d=0~~,\vspace{0.6ex}\\
%%%
a_{p}=c_{p}=0~,~~~p=0,\dots ,~s-1~~,\vspace{0.6ex}\\
%%%
\d=\frac{i(-1)^{s-1}}{s}\left\{\sum_{i=0}^{s-2}~(-1)^{i}\a_{i}-\sum_{i=0}^{s-1}~(-1)^{i}\b_{i}
+(s+1)\sum_{i=0}^{s-1}~(-1)^{i}\g_{i}\right\}~~,\vspace{0.6ex}\n\label{delta}\\
%%%
\kappa=\frac{(-1)^{s-1}}{s}\left\{\sum_{i=0}^{s-2}~(-1)^{i}\a_{i}-\sum_{i=0}^{s-1}~(-1)^{i}\b_{i}
+(s+1)\sum_{i=0}^{s-1}~(-1)^{i}\g_{i}\right\}~~,\vspace{0.6ex}\\
%%%
\xi_{p}=\frac{i}{s}\left\{\vphantom{\frac12}\a_{p}~(s-p)-\b_{p+1}~(p+1)\right\}~~,
~~~p=0,\dots ,~s-2~~,\vspace{0.6ex}\\
%%%
\zeta_{p}=\frac{i(-1)^{p-1}}{s}\left\{\sum_{i=0}^{p-1}~(-1)^{i}\a_{i}-\sum_{i=0}^{p}~(-1)^{i}\a_{i}
+(s+1)\sum_{i=0}^{p}~(-1)^{i}\g_{i}\right\}~,~~~p=0,\dots ,~s-2~~,\vspace{0.6ex}\\
%%%
g_{p}=0~,~~~p=0,\dots ,~s-2~~,
\eea
and the conserved supercurrent multiplet we get is:
\bea{l}
\J_{\a(s)\ad(s)}=
\delta~\pa^{(s)}\Phi~\K_{\Phi}
+\sum_{p=0}^{s-2} \a_{p}~\pa^{(p)}\D\Phi~\pa^{(s-p-1)}\Dd\K_{\Phi}\n\label{bJs4}\\
\hspace{10ex}+\sum_{p=0}^{s-1} \b_{p}~\pa^{(p)}\Phi~\pa^{(s-p-1)}\D\Dd\K_{\Phi}
+\sum_{p=0}^{s-1} \g_{p}~\pa^{(p)}\Phi~\pa^{(s-p-1)}\Dd\D\K_{\Phi}~,\\
\T_{\a(s-1)\ad(s-1)}=\kappa~\pa^{(s-1)}\Phi~\K_{\Phi}
+\sum_{p=0}^{s-2} \zeta_{p}~\pa^{(p)}\Phi~\pa^{(s-p-2)}\Dd\D\K_{\Phi}
+\sum_{p=0}^{s-2} \xi_{p}~\pa^{(p)}\Phi~\pa^{(s-p-2)}\D\Dd\K_{\Phi}~.\n\label{bTs4}
\eea
If we compare \eqref{bJs4} with the corresponding expression for the $s=1$
 case \eqref{s=1:cJ}
we realize that there is a qualitative difference between the two
supercurrents. In \eqref{s=1:cJ},
the $\d$ coefficient was not fixed, whereas for $s\geq2$ the same coefficient
is fixed and given
by \eqref{delta} which was the outcome of the third (3) condition in
\eqref{conserv.cond}. The reason for that is exactly what was mentioned in
\eqref{expand.dK}. The $e\K$ term was independent in $s=1$, whereas it could
be ignored in $s\geq2$. However, if the K\"{a}hler potential was not arbitrary,
but had the special property to remove the third condition in
\eqref{conserv.cond} by making the corresponding term vanish, then by accident
we will be in the same situation as supergravity, where all coefficients in
\eqref{bJs4} are unconstrained. The corresponding term in the conservation
equation that controls this is $\pa^{(s-1)}\Phi~\Dd^2\D\K_{\Phi}$. It will
be interesting to consider potentials $\K$ that make this term vanish
identically. Notice that the free theory $\K=\bar{\Phi}\Phi$ does that.
%*********************************************
\subsection{Reality condition}
%*********************************************
The last step in order to complete our construction is to search for real supercurrents of type
\eqref{bJs4}. Using \eqref{expK}, we can write
\bea{l}
\J_{\a(s)\ad(s)}=
\delta~\pa^{(s)}\Phi~\K_{\Phi}+\sum_{p=0}^{s-2} \a_{p}~\pa^{(p)}\D\Phi~\pa^{(s-p-1)}\left(\vphantom{\frac12}
\Dd\bar{\Phi}~\K_{\Phi\bar{\Phi}}\right)\n\label{expnd.for.real}\\
\hspace{10ex}+\sum_{p=0}^{s-1} \b_{p}~\pa^{(p)}\Phi~\pa^{(s-p-1)}\left(\vphantom{\frac12}
i\pa\bar{\Phi}~\K_{\Phi\bar{\Phi}}+\D\Phi~\Dd\bar{\Phi}~\K_{\Phi\Phi\bar{\Phi}}\right)\\
\hspace{10ex}+\sum_{p=0}^{s-1} \g_{p}~\pa^{(p)}\Phi~\pa^{(s-p-1)}\left(\vphantom{\frac12}
i\pa\Phi~\K_{\Phi\Phi}-\D\Phi~\Dd\bar{\Phi}~\K_{\Phi\Phi\bar{\Phi}}\right)~.
\eea
After distributing the derivatives and collecting similar terms, as we did in \eqref{dc1}, we find
the conditions imposed on coefficients $\a_{p},\b_{p}, \g_{p}$. To simplify this step we split
$\J_{\a(s)\ad(s)}$ into two pieces
\bea{l}
\J_{\a(s)\ad(s)}=\J_{\a(s)\ad(s)}^{(\cancel{\D})}+\J_{\a(s)\ad(s)}^{(\D)}\n
\eea
where $\J_{\a(s)\ad(s)}^{(\cancel{\D})}$ are the contributions to $\J_{\a(s)\ad(s)}$ without spinorial
derivatives and $\J_{\a(s)\ad(s)}^{(\D)}$ are the terms which include spinorial derivatives. This
distinction is useful because terms from one piece can not contribute to the reality of the other,
thus we can examine them separately. Therefore, if we ignore for the moment all the terms in \eqref{expnd.for.real}
with spinorial derivatives we get:
\bea{ll}
\J_{\a(s)\ad(s)}^{(\cancel{\D})}=&~~\underline{\pa^{(s)}\Phi~\left\{\vphantom{\frac12}\d~\K_{\Phi}+
i\g_{0}~\Phi\K_{\Phi\Phi}\right\}+
\pa^{(s)}\bar{\Phi}~\left\{\vphantom{\frac12}i\b_{0}~\Phi\K_{\Phi\bar{\Phi}}\right\}}\\
%%%
&+\underline{\sum_{p=1}^{s-1}~\pa^{(p)}\Phi~\pa^{(s-p)}\Phi~
\left\{\vphantom{\frac12}i\g_{p}~\K_{\Phi\Phi}\right\}}\\
%%%
&+\underline{\sum_{p=1}^{s-1}~\pa^{(p)}\Phi~\pa^{(s-p)}\bar{\Phi}~
\left\{\vphantom{\frac12}i\b_{p}~\K_{\Phi\bar{\Phi}}\right\}}\\
%%%
&+\sum_{p=1}^{s-1}~\pa^{(s-p)}\Phi~\pa^{(p)}\K_{\Phi\Phi}~\left\{\vphantom{\frac12}
i\g_{0}\binom{s-1}{p}\right\}\n\label{r1}
%%%
+\sum_{p=1}^{s-1}~\pa^{(s-p)}\bar{\Phi}~\pa^{(p)}\K_{\Phi\bar{\Phi}}~\left\{\vphantom{\frac12}
i\b_{0}\binom{s-1}{p}\right\}\\
%%%
&+\sum_{p=1}^{s-2}\sum_{q=1}^{s-p-1}~\pa^{(p)}\Phi~\pa^{(s-p-q)}\Phi~\pa^{(q)}\K_{\Phi\Phi}
\left\{\vphantom{\frac12}i\g_{p}\binom{s-p-1}{q}\right\}\\
%%%
&+\underline{\sum_{p=1}^{s-2}\sum_{q=1}^{s-p-1}~\pa^{(p)}\Phi~\pa^{(s-p-q)}\bar{\Phi}~\pa^{(q)}\K_{\Phi\bar{\Phi}}
\left\{\vphantom{\frac12}i\b_{p}\binom{s-p-1}{q}\right\}}~.
\eea
The reality of \eqref{r1} for an arbitrary $\K$ gives the following conditions
\footnote{We have underlined the relevant terms in order for the reader to track the origin
of these conditions.}:
\bea{l}
\d=\g_{0}=\b_{0}=0~~,\n\label{r1r}\vspace{1ex}\\
\g_{p}=0~, ~p=1,2,\dots,s-1~~,\vspace{1ex}\\
\b_{p}=-\b^{*}_{s-p}~, ~p=1,2,\dots,s-1~~,\vspace{1ex}\\
\b_{p}\binom{s-p-1}{q}=-\b^{*}_{s-p-q}\binom{p+q-1}{q}~.
\eea
%
%\begin{equation}\label{r1r}
%\left.\begin{array}{l}
%\d=\g_{0}=\b_{0}=0~~,\vspace{1ex}\\
%\g_{p}=0~, ~p=1,2,\dots,s-1~~,\n\vspace{1ex}\\
%\b_{p}=-\b^{*}_{s-p}~, ~p=1,2,\dots,s-1~~,\vspace{1ex}\\
%\b_{p}\binom{s-p-1}{q}=-\b^{*}_{s-p-q}\binom{p+q-1}{q}
%\end{array} \right\} ~\Rightarrow
%\begin{array}{l}
%\d=0~~,\\
%\g_{p}=0~~,\\
%\b_{p}=ir_{s}\binom{ s-2}{p-1}~,r_{s}\in\mathbb{R}
%\end{array}
%\end{equation}
%
Doing the same for the reality of the terms with spinorial derivatives we get
the constraints:
\bea{l}
\a_{0}=0~~,\vspace{1ex}\\
\b_{s-1}=0~~,\vspace{1ex}\\
\b_{p}=0~,~p=1,2,\dots,s-2~~,\vspace{1ex}\n\label{r2r}\\
\a_{p}=0~,~p=1,2,\dots,s-2~~.
\eea
This is because by using $\b_{0}=\g_{p}=0$, the  terms 
in $\J_{\a(s)\ad(s)}^{(\D)}$ can be written in the following manner:
\bea{ll}
\hspace{-1ex}\J_{\a(s)\ad(s)}^{(\D)}=&~~\underline{\D\Phi~\pa^{(s-1)}\Dd\bar{\Phi}~
\left\{\vphantom{\frac12}\a_{0}~\K_{\Phi\bar{\Phi}}\right\}}\\
%%%%%%
&+\sum_{p=1}^{s-2}~\pa^{(p)}\D\Phi~\pa^{(s-p-1)}\Dd\bar{\Phi}~
\left\{\vphantom{\frac12}\a_{p}~\K_{\Phi\bar{\Phi}}\right\}
%%%%%%
+\sum_{p=1}^{s-2}~\D\Phi~\pa^{(s-p-1)}\Dd\bar{\Phi}~\pa^{(p)}\K_{\Phi\bar{\Phi}}
\left\{\vphantom{\frac12}\a_{0}\binom{s-1}{p}\right\}\\
%%%%%%
&+\sum_{p=1}^{s-2}~\pa^{(p)}\Phi~\D\Phi~\pa^{(s-p-1)}\Dd\bar{\Phi}
\left\{\vphantom{\frac12}\b_{p}~\K_{\Phi\Phi\bar{\Phi}}\right\}\\
&+\sum_{p=1}^{s-2}\sum_{q=1}^{s-p-2}~\pa^{(p)}\Phi~\D\Phi~\pa^{(s-p-q-1)}\Dd\bar{\Phi}~
\pa^{(q)}\K_{\Phi\Phi\bar{\Phi}}\left\{\vphantom{\frac12}\b_{p}\binom{s-p-1}{q}\right\}\\
%%%%%%
&+\underline{\sum_{p=1}^{s-2}~\pa^{(s-p-1)}\D\Phi~\Dd\bar{\Phi}~\pa^{(p)}\K_{\Phi\bar{\Phi}}
\left\{\vphantom{\frac12}\a_{s-p-1}\right\}}
%%%%%%
+\sum_{p=1}^{s-2}~\pa^{(p)}\Phi~\pa^{(s-p-1)}\D\Phi~\Dd\bar{\Phi}~
\left\{\vphantom{\frac12}\b_{p}~\K_{\Phi\Phi\bar{\Phi}}\right\}\\
%%%%%%
&+\sum_{p=1}^{s-2}\sum_{q=1}^{s-p-2}~\pa^{(p)}\Phi~\pa^{(s-p-q-1)}\D\Phi~\Dd\bar{\Phi}~
\pa^{(q)}\K_{\Phi\Phi\bar{\Phi}}
\left\{\vphantom{\frac12}\b_{p}\binom{s-p-1}{q}\right\}\n\label{r2}\\
%%%%%%
&+\D\Phi~\Dd\bar{\Phi}~\pa^{(s-1)}\K_{\Phi\bar{\Phi}}
\left\{\vphantom{\frac12}\a_{0}\right\}+\underline{\pa^{(s-1)}\Phi~\D\Phi~\Dd\bar{\Phi}
\left\{\vphantom{\frac12}\b_{s-1}~\K_{\Phi\Phi\bar{\Phi}}\right\}}\\
%%%%%%
&+\underline{\sum_{p=1}^{s-2}~\pa^{(p)}\Phi~\D\Phi~\Dd\bar{\Phi}~\pa^{(s-p-1)}\K_{\Phi\Phi\bar{\Phi}}
\left\{\vphantom{\frac12}\b_{p}\right\}}\\
%%%%%%
&+\sum_{p=1}^{s-2}\sum_{q=1}^{s-p-2}~\pa^{(p)}\D\Phi~\pa^{(s-p-q-1)}\Dd\bar{\Phi}~
\pa^{(q)}\K_{\Phi\bar{\Phi}}\left\{\vphantom{\frac12}\a_{p}\binom{s-p-1}{q}\right\}\\
%%%%%%
&+\sum_{p=1}^{s-2}\sum_{q=1}^{s-p-2}~\pa^{(p)}\Phi~\pa^{(q)}\D\Phi~\pa^{(s-p-q-1)}\Dd\bar{\Phi}~
\left\{\vphantom{\frac12}\b_{p}\binom{s-p-1}{q}~\K_{\Phi\Phi\bar{\Phi}}\right\}\\
%%%%%%
&+\sum_{p=1}^{s-2}\sum_{q=1}^{s-p-2}\sum_{r=1}^{s-p-q-2}\pa^{(p)}\Phi~\pa^{(r)}\D\Phi~
\pa^{(s-p-q-r-1)}\Dd\bar{\Phi}~\pa^{(q)}\K_{\Phi\Phi\bar{\Phi}}
\left\{\vphantom{\frac12}\b_{p}\binom{s-p-1}{q}\binom{s-p-q-1}{r}\right\}.
\eea
The conclusion is that, in order to get a real, higher spin supercurrent all coefficients
$\a_{p},\b_{p},\g_{p}$ must vanish.
Therefore, there is no non-trivial solution for arbitrary K\"{a}hler potential $\K$.

However, an interesting question one can ask is whether there is a special K\"{a}hler potential $\K^{s}$
such that we can construct non-trivial real, higher spin supercurrents. After all, we have seen this
behavior in the coupling with the vector supermultiplet, where the K\"{a}hler potential must have a
$U(1)$ symmetry. Therefore, one can imagine that if the
special property of $\K^{s}$ is a realization of \emph{higher spin symmetry} then maybe the  higher spin
supercurrent  exist. Going back to \eqref{r2} and examining the terms responsible for the vanishing of
$\b_{p}$ and $\a_{p}$ we find that a necessary condition for $\K^{s}$ is:
\bea{l}
\K^{s}_{\Phi\Phi\bar{\Phi}}=0\n\label{sK}~.
\eea
This is equivalent to $\K^{s}_{\Phi\bar{\Phi}}=\textit{constant}$ which holds true only for the free theory
$\K^{s}=\bar{\Phi}\Phi$. Furthermore, this condition is consistent because for the free theory the term
$\pa^{(s-1)}\Phi~\Dd^2\D\K_{\Phi}$ leading to the fixing of $\delta$ in \eqref{delta} vanishes, hence there
is no incompatibility between the non-trivial values of the parameters and conservation equation.
Additionally, the quantity $Z_{\a(s-1)\ad(s-1)}$ \eqref{extra.freedom} becomes identically zero for the
free theory. So the parameters we removed, such as $\a_{s-1}$ become relevant now. All these
\emph{accidents}
take place only if \eqref{sK} is true and for that case we recover the results of
\cite{Superspace1,Superspace2}.

%%%%%%%%%%%%%%%%%%%%%%%%%%%%%%%%%%%%%%%%%%%%%%%%%
\section[Turn on chiral superpotential W]{Turn on the chiral superpotential $\W$}
\label{sec:W}
%%%%%%%%%%%%%%%%%%%%%%%%%%%%%%%%%%%%%%%%%%%%%%%%%
In this section, we turn back on the chiral superpotential $\W(\Phi)$ in order to
study its contribution to the higher spin supercurrent multiplet, when that
is possible. We must keep in mind that $\W$ is a chiral superfield
($\Dd_{\ad}\W=0$) and the on-shell equation of motion now becomes
$\Dd^2\K_{\Phi}=\W_{\Phi}$.
%*********************************************
\subsection{For vector supermultiplet}
%*********************************************
We start with supercurrent \eqref{s=0:J} and we modify it in order to include
the chiral superpotential information while preserving its reality:
\bea{l}
\J=\Phi\K_{\Phi}+c~\Lambda\W_{\Phi}+c^*~\bar{\Lambda}
\mathcal{\bar{W}}_{\bar{\Phi}}\n
\eea
where $\Lambda$ is the prepotential of the chiral superfield,
$\Phi=\Dd^2\Lambda$. It is straight forward to prove that the conservation
equation of $\J$
\bea{l}
\Dd^2\J=(1+c)~\Phi\W_{\Phi}
+c^*~\Dd^2\bar{\Lambda}\mathcal{\bar{W}}_{\bar{\Phi}}
+c^*~\Dd^{\bd}\bar{\Lambda}\Dd_{\bd}\mathcal{\bar{W}}_{\bar{\Phi}}
+c^*~\bar{\Lambda}\Dd^2\mathcal{\bar{W}}_{\bar{\Phi}}=0\n
\eea
can not be satisfied for any value of $c$. Hence, the conclusion
is that in the presence of any chiral superpotential $\W$, there can be no
conserved supercurrent no matter what the K\"{a}hler potential is, even
for the free theory. This can be understood as the fact that
the presence of $\W$ breaks the global $U(1)$ symmetry of section
\ref{sec:vector}.
%*********************************************
\subsection{For supergravity}
%*********************************************
For the supercurrent multiplet (\ref{J1},\ref{T1}) that generates interactions
with the supergravity supermultiplet we consider the following modification
terms
\bea{l}
\J_{\a\ad}=\D_{\a}\Phi~\Dd_{\ad}\K_{\Phi}+c_1~\D_{\a}\Dd_{\ad}\Lambda~\F
+c_2~\Dd_{\ad}\Lambda~\D_{\a}\F+c_3~\Lambda~\Dd_{\ad}\D_{\a}\F
+c_4~\Dd_{\ad}\D_{\a}\Lambda~\F\n\\
%%%%
\hspace{19.9ex}
-c^{*}_1~\Dd_{\ad}\D_{\a}\bar{\Lambda}~\mathcal{\bar{F}}
-c^{*}_2~\D_{\a}\bar{\Lambda}~\Dd_{\ad}\mathcal{\bar{F}}
-c^{*}_3~\bar{\Lambda}~\D_{\a}\Dd_{\ad}\mathcal{\bar{F}}
-c^{*}_4~\D_{\a}\Dd_{\ad}\bar{\Lambda}~\mathcal{\bar{F}}~,\\
\T=-\K+d_1~\Lambda\F+d_2~\bar{\Lambda}\mathcal{\bar{F}}~.\n
\eea
In the above terms, $\F=\F(\Phi)$ is a chiral superfield and a holomorphic
function of $\Phi$ defined as
\bea{l}
\W(\Phi)=\Phi\F(\Phi)~.\n
\eea
This definition holds for any chiral superpotential because its Taylor
expansion does not include the constant term due to \eqref{eqclW}.
Additionally, it
relates $\W_{\Phi}$ which appears in the equation of motion
with $\F$ in the following manner:
\bea{l}
\W_{\Phi}=\F+\Phi\F_{\Phi}~.\n
\eea
Imposing the conservation equation \eqref{s=1:ce} we get a non-trivial
solution
\bea{ll}
c_1=-1~,~& d_1=1~~,\\
c_2=1~,~& d_2=2~~,\n\\
c_3=0~,~& c_4=0~.
\eea
Thus, the supercurrent multiplet (\ref{J1},\ref{T1}) can be generalized to
include an arbitrary chiral superpotential
\bea{l}
\J_{\a\ad}=\D_{\a}\Phi~\Dd_{\ad}\K_{\Phi}-\D_{\a}\Dd_{\ad}(\Lambda\F)
+\Dd_{\ad}\D_{\a}(\bar{\Lambda}\mathcal{\bar{F}})~,\n\\
\T=-\K+\Lambda\F+2\bar{\Lambda}\mathcal{\bar{F}}~.\n
\eea
%***********************************************
\subsection{For higher spin supermultiplet}
%***********************************************
In section \ref{sec:hs} we showed that the construction of the
higher spin supercurrent multiplet is possible only for the free
theory, $\K=\bar{\Phi}\Phi$. For that case \cite{Superspace1,Superspace2},
there is the \emph{minimal} multiplet
\bea{l}
\J^{\textit{free}}_{\a(s)\ad(s)}=
c(-i)^s\sum_{p=0}^{s}(-1)^{p}\binom{s}{p}^2~\pa^{(p)}\Phi~
\pa^{(s-p)}\bar{\Phi}\sn\\
\hspace{10ex}+ic(-i)^s\sum_{p=0}^{s-1}(-1)^{p}\binom{s}{p}^2\frac{s-p}{p+1}~
\pa^{(p)}\D\Phi~\pa^{(s-p-1)}\Dd\bar{\Phi}~,\\
\T^{free}_{\a(s-1)\ad(s-1)}=0~,\sn
\eea
where $c$ is a real proportionality constant (it may depend on the value of $s$). Hence, the consideration of
contributions due to the presence of a chiral superpotential $\W$ must take
place in the same configuration. The most general ansatz for the $\W$
generated terms are:
\bea{l}
\J^{\W}_{\a(s)\ad(s)}=~~\sum_{p=0}^{s-1}\g_{p}~\pa^{(p)}\D\Dd\Lambda~
\pa^{(s-p-1)}\W_{\Phi}+\sum_{p=0}^{s-1}\d_{p}~\pa^{(p)}\Dd\Lambda~
\pa^{(s-p-1)}\D\W_{\Phi}\n\label{Jw}\vspace{0ex}\\
%%%%%%%%%
\hspace{11ex}
-\sum_{p=0}^{s-1}\g^{*}_{p}~\pa^{(p)}\Dd\D\bar{\Lambda}~
\pa^{(s-p-1)}\mathcal{\bar{W}}_{\bar{\Phi}}
-\sum_{p=0}^{s-1}\d^{*}_{p}~\pa^{(p)}\D\bar{\Lambda}~
\pa^{(s-p-1)}\Dd\mathcal{\bar{W}}_{\bar{\Phi}}~~,\vspace{0ex}\\
%%%%%%%%%%%
\T^{\W}_{\a(s-1)\ad(s-1)}=\sum_{p=0}^{s-1}\zeta_{p}~\pa^{(p)}\Lambda~
\pa^{(s-p-1)}\W_{\Phi}
%%%%%
+\sum_{p=0}^{s-2}\s_{p}~\pa^{(p)}\Dd\D\Lambda~\pa^{(s-p-2)}\W_{\Phi}
%%%%%
+\sum_{p=0}^{s-1}\xi_{p}~\pa^{(p)}\bar{\Lambda}~
\pa^{(s-p-1)}\mathcal{\bar{W}}_{\bar{\Phi}}~,~~~~~~\n\label{Tw}
\eea
and the conservation equation they must satisfy is
\bea{l}
\Dd^{\ad_s}\J^{\textit{free}}_{\a(s)\ad(s)}+
\Dd^{\ad_s}\J^{\W}_{\a(s)\ad(s)}-
\frac{1}{s!}\Dd^2\D_{(\a_s}\T^{\W}_{\a(s-1))\ad(s-1)}=0~.\n\label{cew}
\eea
Substituting (\ref{Jw},\ref{Tw}) in \eqref{cew}, the cancellation of the
$\Lambda$ and $\bar{\Lambda}$ dependent terms gives:
\bea{l}\n
\d_{p}=-\g_{p}~~,~~p=0,1,\dots,s-1~~,\sn\\
\xi_{p}=-\tfrac{s+1}{s}\g_{p}^{*}~~,~~p=0,1,\dots,s-1~~,\sn\\
\zeta_{0}=-\tfrac{1}{s}\g_{0}~~,\sn\\
\zeta_{p}=-\tfrac{p+1}{s}\g_{p}+\tfrac{s-p}{s}\g_{p-1}
~~,~~p=1,\dots,s-1~~,\sn\label{zeta}\\
i\s_{0}=\tfrac{1}{s}\g_{1}-\tfrac{s-1}{s}\g_{0}~~,\sn\\
i\s_{s-2}=\tfrac{s-1}{s}\g_{s-1}-\tfrac{1}{s}\g_{s-2}~~,\sn\\
i\s_{p}+i\s_{p-1}=\tfrac{p+1}{s}\g_{p+1}-\tfrac{s-2p-1}{s}\g_{p}
-\tfrac{s-p}{s}\g_{p-1}~~,~~p=0,1,\dots,s-1~~.\sn
\eea
These are exactly the conditions found in \cite{Superspace2}. The cancellation
of the $\Phi$ dependent terms gives:
\bea{l}
0=\left\{\vphantom{\frac12}c(i)^{s+1}(s+1)-\g_{s-1}\right\}~\pa^{(s-1)}\D\Phi~
\W_{\Phi}
+\left\{\vphantom{\frac12}c(-i)^{s-1}-\g_{0}\right\}~\Phi~
\pa^{(s-1)}\D\W_{\Phi}\\
%%%%%%
\hspace{3ex}+\sum_{p=0}^{s-2}\left\{\vphantom{\frac12} c(-i)^{s+1}(-1)^{p}~\frac{s+1}{s}
\binom{s}{p}^2\frac{s-p}{p+1}-\frac{p+1}{s}
\left(\vphantom{\frac12}  \g_{p+1}+\g_{p}\right)
\right\}~\pa^{(p)}\D\Phi~\pa^{(s-p-1)}\W_{\Phi}\n\label{fix-g}\\
%%%%%%%
\hspace{3ex}+\sum_{p=0}^{s-2}\frac{s-p-1}{p+1}
\left\{\vphantom{\frac12} c(-i)^{s+1}(-1)^{p}~\frac{s+1}{s}
\binom{s}{p}^2\frac{s-p}{p+1}-\frac{p+1}{s}
\left(\vphantom{\frac12}  \g_{p+1}+\g_{p}\right)
\right\}
~\pa^{(p+1)}\Phi~\pa^{(s-p-2)}\D\W_{\Phi}~.
\eea
Therefore, for an arbitrary chiral superpotential we get the constraints
\bea{l}\n
\g_{s-1}=c(i)^{s+1}(s+1)~,\sn\label{gs-1}\\
\g_{0}=c(-i)^{s-1}~,\sn\\
\g_{p+1}+\g_{p}=c(-i)^{s+1}(-1)^{p}(s+1)\binom{s}{p}^2\frac{s-p}{(p+1)^2}~,
~~p=0,1,\dots,s-2\sn\label{gp}~.
\eea
One can easily check that the above equations are not consistent
with each other. For example, starting with $\g_{0}$
and using the recursive equation \eqref{gp} we reach to a different value of
$\g_{s-1}$ than the one required.
The conclusion is
that for an arbitrary chiral superpotential $\W$, there is no higher spin
supercurrent multiplet.

 Nevertheless, careful
examination of \eqref{fix-g} reveals two exceptions exist for some special
superpotential $\W^{s}$. The first one is for
\bea{l}
\W^{s}\sim\Phi^2.\n\label{mPhi}
\eea
In this case, the various terms of \eqref{fix-g} are no longer independent
and they can be combined, resulting to a different set of constraints for the
$\g_{p}$ parameters. Of course, \eqref{mPhi} corresponds to the mass term of the
chiral superfield and equation \eqref{fix-g} will lead to the analysis of
\cite{Superspace2} were the mass term contributions to the higher spin
supercurrent and supertrace were calculated. The main result was that only the
odd values of $s$ ($s=2l+1$) will lead to consistent, interactions.
Additionally, a second exception exist and corresponds to a linear superpotential
\bea{l}
\W^{s}=f\Phi~.\n
\eea
In that case all the terms of \eqref{fix-g} with derivatives acting on
$\W_{\Phi}$ will vanish and we get only condition \eqref{gs-1}. The
corresponding supercurrent multiplet is:
\bea{l}
\J_{\a(s)\ad(s)}=\J^{\textit{free}}_{\a(s)\ad(s)}+fc(s+1)(i)^{s+1}~
\pa^{(s-1)}\D\Dd\Lambda-\bar{f}c(s+1)(-i)^{s+1}~\pa^{(s-1)}\Dd\D\bar{\Lambda}
~,\n\\
\T_{\a(s-1)\ad(s-1)}=-fc~(s+1)(i)^{s+1}~\pa^{(s-1)}\Lambda
+fc~\tfrac{s^2-1}{s}~(i)^{s}~\pa^{(s-2)}\Dd\D\Lambda
-\bar{f}c~\tfrac{(s+1)^2}{s}~(-i)^{s+1}~\pa^{(s-1)}\bar{\Lambda}~.~~~~\n
\eea
and it exists for all values of $s$.
%%%%%%%%%%%%%%%%%%%%%%%%%%%%%%%%%%%%%%%%%%%%%%%%%
\section{Summary and Discussion}
\label{sec:discussion}
%%%%%%%%%%%%%%%%%%%%%%%%%%%%%%%%%%%%%%%%%%%%%%%%%
In this paper, we have investigated the interactions of higher spin gauge fields
with matter theory fields independently from the ability to have a properly
define $\mathcal{S}$-matrix. We considered an interacting matter theory described
by a chiral superfield $\Phi$ with its dynamics been described an arbitrary
K\"{a}hler potential $\K(\Phi,\bar{\Phi})$ and a chiral
superpotential $\W(\Phi)$. The arbitrariness of both $\K$ and $\W$ 
allows us to parametrize some very complicated, strong interactions which can disrupt the
conventional definition of free \emph{in} and \emph{out} states and thus evading the
consequences of the Coleman-Mandula theorem. We have also assumed that the first order
interactions of the matter theory with higher spin supermultiplets of type 
($s+1,s+1/2$), if they exist, are generated by a higher spin supercurrent 
multiplet $(\J_{\a(s)\ad(s)},T_{\a(s-1)\ad(s-1)})$
defined by a real higher spin supercurrent $\J_{\a(s)\ad(s)}$ and a higher spin supertrace
$\T_{\a(s-1)\ad(s-1)}$
\bea{l}
S_1\sim\int
d^8z\left\{\vphantom{\frac12}
H^{\a(s)\ad(s)}\J_{\a(s)\ad(s)}
+\chi^{\a(s)\ad(s-1)}\D_{\a_s}\T_{\a(s-1)\ad(s-1)}
+\bar{\chi}^{\a(s-1)\ad(s)}\Dd_{\ad_s}\mathcal{\bar{T}}_{\a(s-1)\ad(s-1)}\right\}
\eea
and on-shell satisfy the conservation equation
\bea{l}
\Dd^{\ad_{s}}\J_{\a(s)\ad(s)}=\tfrac{1}{s!}\Dd^2\D_{(\a_{s}}\T_{\a(s-1))\ad(s-1)}~.
\eea
The results we find are:
\begin{enumerate}
\item[(\emph{i})] For $s=0$ (vector supermultiplet) there are no interactions
unless the K\"{a}hler potential can be written as a function of the product
$\bar{\Phi}\Phi$ ~[$\K=\K(\bar{\Phi}\Phi)$] and the chiral superpotential
vanishes ~[$\W(\Phi)=0$]. For these cases, the
supercurrent is
\bea{l}
\J=\Phi\K_{\Phi}
\eea
and satisfies the conservation equation $\Dd^2\J=0$. The  constraints in
$\K$ and $\W$ can be understood as the global $U(1)$ symmetry
requirement for the gauging procedure which will generate the interactions
with the vector supermultiplet.
%%%%%
\item[(\emph{ii})] For $s=1$ (non-minimal supergravity supermultiplet)
the expectation is that we should always be able to find interactions.
This was verified by our approach because for any $\K$ and $\W$ we can construct the
following supercurrent multiplet
\bea{l}
\J_{\a\ad}=\D_{\a}\Phi~\Dd_{\ad}\K_{\Phi}-\D_{\a}\Dd_{\ad}(\Lambda\F)
+\Dd_{\ad}\D_{\a}(\bar{\Lambda}\mathcal{\bar{F}})~,\\
\T=-\K+\Lambda\F+2\bar{\Lambda}\mathcal{\bar{F}}~,
\eea
where $\Phi=\Dd^2\Lambda$ and $\W=\Phi\F$.
Furthermore, this result is consistent with the results of
\cite{Magro}.
%%%
\item[(\emph{iii})] For $s\geq2$ (higher spin supermultiplets) there is
a severe constraining of both $\K$ and $\W$. For almost any
K\"{a}hler potential and chiral superpotential there are no interactions with higher spin
supermultiplets. In other words, one can not construct a non-trivial higher
spin supercurrent $\J_{\a(s)\ad(s)}$ and supertrace $\T_{\a(s-1)\ad(s-1)}$.
However there are three exceptions:
\begin{enumerate}
\item[1.)] $\K=\bar{\Phi}\Phi$~,~$\W=0$~~: Free, massless, chiral superfield
\item[2.)] $\K=\bar{\Phi}\Phi$~,~$\W=f\Phi$~~: Free, chiral superfield with linear superpotential
\item[3.)] $\K=\bar{\Phi}\Phi$~,~$\W=m\Phi^2$~~: Free, massive, chiral superfield
\end{enumerate}
These exceptions are consistent with various known no-go theorems
such as the Coleman-Mandula theorem \cite{nogo3} and its extensions to
supersymmetry \cite{nogo4}. For exceptions (1) and (3) the corresponding
supercurrent multiplet have been constructed in \cite{Superspace1,Superspace2,Superspace3}.
To this list we add the supercurrent multiplet for exception (2)
\bea{l}
\J_{\a(s)\ad(s)}=\J^{\textit{free}}_{\a(s)\ad(s)}+fc(s+1)(i)^{s+1}~
\pa^{(s-1)}\D\Dd\Lambda-\bar{f}c(s+1)(-i)^{s+1}~\pa^{(s-1)}\Dd\D\bar{\Lambda}
~,\\
\T_{\a(s-1)\ad(s-1)}=-fc~(s+1)(i)^{s+1}~\pa^{(s-1)}\Lambda
+fc~\tfrac{s^2-1}{s}~(i)^{s}~\pa^{(s-2)}\Dd\D\Lambda
-\bar{f}c~\tfrac{(s+1)^2}{s}~(-i)^{s+1}~\pa^{(s-1)}\bar{\Lambda}~.
\eea
\end{enumerate}
%%%%%%%%%%%%%%%%%%%%%%%%%%%%%%%%%%%%%%%%%%%%%%%%%
{\bf Acknowledgments}\\[.1in] \indent
The research of I.\ L.\ B.\ was supported in parts by
Russian Ministry of Education and Science, project No. 3.1386.2017.
He is also grateful to RFBR grant, project No. 18-02-00153 for
partial support. The research of S.\ J.\ G.\ is supported by the 
endowment of the Ford Foundation Professorship of Physics at 
Brown University. The work of K.\ K.\ was supported by the grant
P201/12/G028 of the Grant agency of Czech Republic.
K.\ K.\ is thankful to Rikard von Unge and Linus Wulff for useful
discussions.
%%%%%%%%%%%%%%%%%%%%%%%%%%%%%%%%%%%%%%%%%%%%%%%%%

%%%%%%%%%%%%%%%%%%%%%%%%%%%%%%%%%%%%

\end{document}